\documentstyle[mnras_cite,psfig]{mn}
\def\gsim{~\rlap{$>$}{\lower 1.0ex\hbox{$\sim$}}}
\def\lsim{~\rlap{$<$}{\lower 1.0ex\hbox{$\sim$}}}
\def\d{{\rm d}}

\begin{document} 

\title{Galaxy Voids in Cold Dark Matter Universes}  
\author[A.~J.~Benson, Fiona~Hoyle, Fernando~Torres  \& Michael~S.~Vogeley]{A.~J.~Benson$^1$, Fiona~Hoyle$^2$, Fernando~Torres$^3$, \& Michael~S.~Vogeley$^2$ \\ 
1. California Institute of Technology, MC 105-24, Pasadena, CA 91125,
U.S.A. (e-mail: abenson@astro.caltech.edu) \\
2. Department of Physics, Drexel University, Philadelphia, PA 19104, U.S.A. (e-mail: hoyle@venus.physics.drexel.edu,vogeley@drexel.edu) \\
3. Department of Physics, University of California, Davis, 1 Shields Avenue, Davis, CA 95616, U.S.A. (nando823@aol.com)}

\maketitle

\begin{abstract}
We present predictions for numerous statistics related to the presence
of voids in the distribution of galaxies in a cold dark matter model
of structure formation using a semi-analytic model of galaxy
formation. Our study is able to probe galaxies with masses as low as
$10^9h^{-1}M_\odot$ corresponding to absolute magnitudes of $M_{\rm
b_J}-5\log h=-18.1$ and $M_{\rm r}-5\log h=-18.7$. We quantify the
void and underdense probability functions, distributions of nearest
neighbour distances and void sizes and compute the density profiles of
voids. These results are contrasted with the expectations for dark
matter (and the difference examined in terms of the galaxy/dark matter
biasing relation) and are compared to analytic predictions and
observational data where available. The predicted void probability
functions are consistent with those measured from the Center for
Astrophysics redshift surveys given the rather large uncertainties in
this relatively small (for studies of voids) observational sample. The
size of the observational sample is too small to probe the bias
between galaxies and dark matter that we predict. We also examine the
predicted properties of galaxies living within voids and contrast
these with the general galaxy population. Our predictions are aimed at
forthcoming large galaxy redshift surveys which should for the first
time provide statistically accurate measures of the void population.\\

\end{abstract}

\section{Introduction}
One of the most striking features of galaxy redshift surveys is the
presence of large regions of space that are nearly devoid of
galaxies. These ``voids'' are thought to form from the most underdense
regions of the initial density field, although other suggestions exist
such as cosmic explosions \cite{ocow81} or first order phase
transitions \cite{amendola99}. While voids have been seen in redshift
surveys since the late 1970's \cite{greg78,kirshner81,geller89}, due
to their large size and low number density it is only with the advent
of the Two-degree Field Galaxy Redshift Survey (2dFGRS) and Sloan
Digital Sky Survey (SDSS) that the statistical properties of voids
will be quantified in a meaningful way.

It is well known that the cold dark matter (CDM) cosmogony produces
voids in the distribution of dark matter (i.e. highly underdense
regions which are not, however, completely empty of dark matter), and
the properties of these voids are well studied
\cite{einasto91,ghigna94,little94,vogeley91,vogeley94,ghigna96,kns97,muller00,armu,sheth02}.
The properties of voids, and the galaxies which live within them have
been proposed as a strong test of the CDM scenario
\cite{peebles01}. However, there have been very few theoretical
studies of voids in the \emph{galaxy} distribution expected in CDM
(with the notable exception of \pcite{mathis02}), primarily due to the
lack of a detailed, physical model for galaxy formation in the
past. It is known both theoretically and observationally that at least
some galaxies are biased tracers of the dark matter (Davis \& Geller
1976; for recent results see Hoyle et al. 1999, Benson et al. 2000a),
even though $L_*$ galaxies may be close to unbiased, as shown in the
2dFGRS by \scite{lverde}. Recent studies of the dependence
of clustering amplitude on galaxy luminosity and morphology in the
SDSS \cite{zehavi02,zehavi03} and the 2dFGRS
\cite{norberg01a,norberg01} show a substantial increase with
luminosity of the correlation function amplitude which is only partly
induced by the variations in the morphological mix of
galaxies. Consequently, we cannot expect voids in the distributions of
galaxies and dark matter to have the same statistical properties. In
fact we will show that they can be quite different.

In this work we aim to remedy this deficiency by presenting
predictions for the simplest and most useful statistical quantifiers
of galaxy voids in a CDM universe using a combination of N-body
simulations of dark matter and semi-analytical modelling of galaxy
formation. This technique has been demonstrated to naturally explain
several aspects of galaxy clustering --- such as the near power-law
shape of the galaxy correlation function and the dependence of galaxy
clustering on luminosity \cite{kauffmann99,cluster1,cluster3}. Such an
approach is currently the only means to make physically realistic
predictions for galaxy voids (the only competitive method for
modelling galaxy formation --- direct hydrodynamical simulation ---
cannot currently be applied to sufficiently large volumes of the
Universe with the desired resolution). We will make predictions
specifically aimed at the 2dFGRS and SDSS surveys, and will contrast
our results with previous analytical and numerical studies.

The remainder of this paper is arranged as follows. In
\S\ref{sec:analysis} we describe the construction of galaxy catalogues
and the details of the analysis which we apply to them. In
\S\ref{sec:statistics}, \S\ref{sec:props} and \S\ref{sec:vgals} we
present results for a variety of statistical properties of voids and
the galaxies within them. Finally, in \S\ref{sec:discuss} we present
our conclusions.

\section{Analysis}
\label{sec:analysis}

Catalogues of mock galaxies are constructed using the techniques of
\scite{cluster3}, to which the reader is referred for full
details. Briefly, we use the semi-analytic model of \scite{cole2000}
with extensions due to \scite{benson02} to populate dark matter halos
located in N-body simulations with galaxies. In this particular
approach, dark matter halos are located in the simulation using the
friends-of-friends technique together with an energy criterion to
ensure the halos represent physical, bound objects. We retain halos
containing ten or more particles. The semi-analytic model is used to
predict the properties of galaxies occupying each halo by accounting
for the rate at which gas can cool and turn into stars, galaxy-galaxy
mergers and feedback from supernovae. Galaxies are then assigned a
position (both in real and redshift-space) within the halo, resulting
in a three dimensional distribution of galaxies.

We use the same two N-body simulations as \scite{cluster3}, which we
refer to as the ``GIF'' and ``$512^3$'', which have cosmological
parameters
$(\Omega_0,\Lambda_0,h,\sigma_8,\Gamma)=(0.3,0.7,0.7,0.9,0.21)$\footnote{We
define Hubble's constant as $H_0=100 h$km/s/Mpc.} and particle masses
of $1.4\times 10^{10}$ and $6.8\times 10^{10}h^{-1}M_\odot$
respectively. We also make use of a third simulation, which we refer
to as ``GIF-II'', which has the exact same parameters as the GIF
simulation, but with a different realization of the initial
conditions. This GIF-II simulation will therefore provide a means to
quantify uncertainties in the results from the GIF simulation due to
its limited volume. The two simulations differ in mass resolution and
volume as detailed in Table~\ref{tb:sims}, where we also list the
faintest absolute magnitudes in ${\rm b_J}$ and r-bands (appropriate
to the 2dFGRS and SDSS respectively) to which our mock galaxy
catalogues are complete in each simulation, along with the
characteristic magnitude $M_*$ measured in these two surveys. (For the
$512^3$ simulation these limits are comparable to $M_*$, while for the
GIF simulation they are almost two magnitudes fainter than $M_*$.) At
fainter magnitudes the limited resolution of the simulation means that
some galaxies are missed. These magnitudes are relatively bright,
perhaps significantly brighter than we might expect to be typical of
``void galaxies''. Higher resolution simulations, allowing us to probe
to fainter magnitudes, would clearly provide stronger tests of the
void phenomenon in CDM models. Note that with our scheme for
constructing mock galaxy catalogues, morphological evolution is
correctly tracked for all galaxies brighter than this luminosity
limit. The GIF simulation allows us to examine somewhat fainter
galaxies than the $512^3$ but is of limited usefulness for studying
voids because of its relatively small volume (voids have very low
number densities). The $512^3$ simulation on the other hand is of
sufficiently large volume to provide a statistically useful sample of
voids. We will construct galaxy catalogues for a variety of absolute
magnitude limits. For each galaxy catalogue we construct we create a
``dark matter'' catalogue with the same number density of points by
randomly selecting dark matter particles from the N-body
simulations. These dark matter catalogues will allow us to examine
differences between the properties of voids in the dark matter and
galaxy distributions (i.e. the ``bias'' of the galaxy
distribution). Voids in an unbiased population of galaxies would have
statistical properties identical to those in our dark matter
catalogues. All catalogues use distributions of galaxies
or dark matter in redshift space (i.e. we shift each galaxy in the
catalogue a distance $v_{\rm x}/{\rm H_0}$, where $v_{\rm x}$ is the
peculiar velocity of the galaxy, along the x-direction to account for
redshift space distortions).

\begin{table*}
\caption{Properties of the two N-body simulations used in this work
and of the mock catalogues constructed from them. Limiting magnitudes
indicate the faintest galaxies for which our samples will be complete,
due to the limited resolution of each simulation.}
\label{tb:sims}
\begin{tabular}{ccccccc}
\hline
 & & & \multicolumn{2}{c}{Limiting magnitude} & \multicolumn{2}{c}{Characteristic magnitude} \\
Simulation & Particle mass ($h^{-1}M_\odot$) & Volume ($h^3$ Mpc$^{-3}$) & $M_{\rm {b_J}}-5\log h$ & $M_{\rm r}-5\log h$ & $M^*_{\rm {b_J}}-5\log h$ & $M^*_{\rm r}-5\log h$ \\
\hline
GIF & $1.4\times 10^{10}$ & $2.2 \times 10^6$ & $-18.1$ & $-18.7$ & $-19.79$ & $-20.83$ \\
$512^3$ & $6.8\times 10^{10}$ & $1.1 \times 10^8$ & $-19.8$ & $-20.6$ & $-19.79$ & $-20.83$ \\
\hline
\end{tabular}
\end{table*}

\begin{figure*}
\begin{tabular}{cc}
\psfig{file=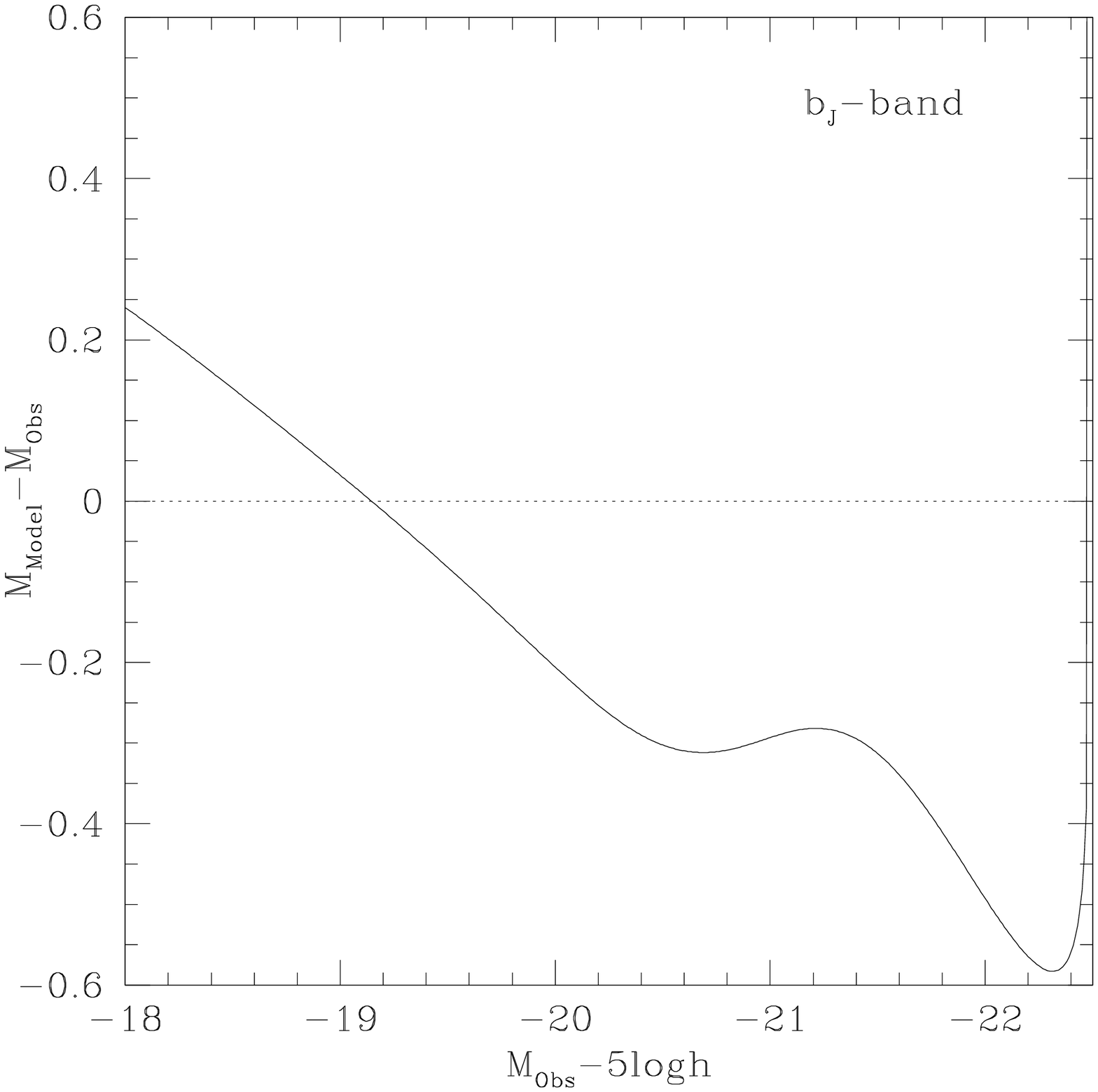,width=80mm} & \psfig{file=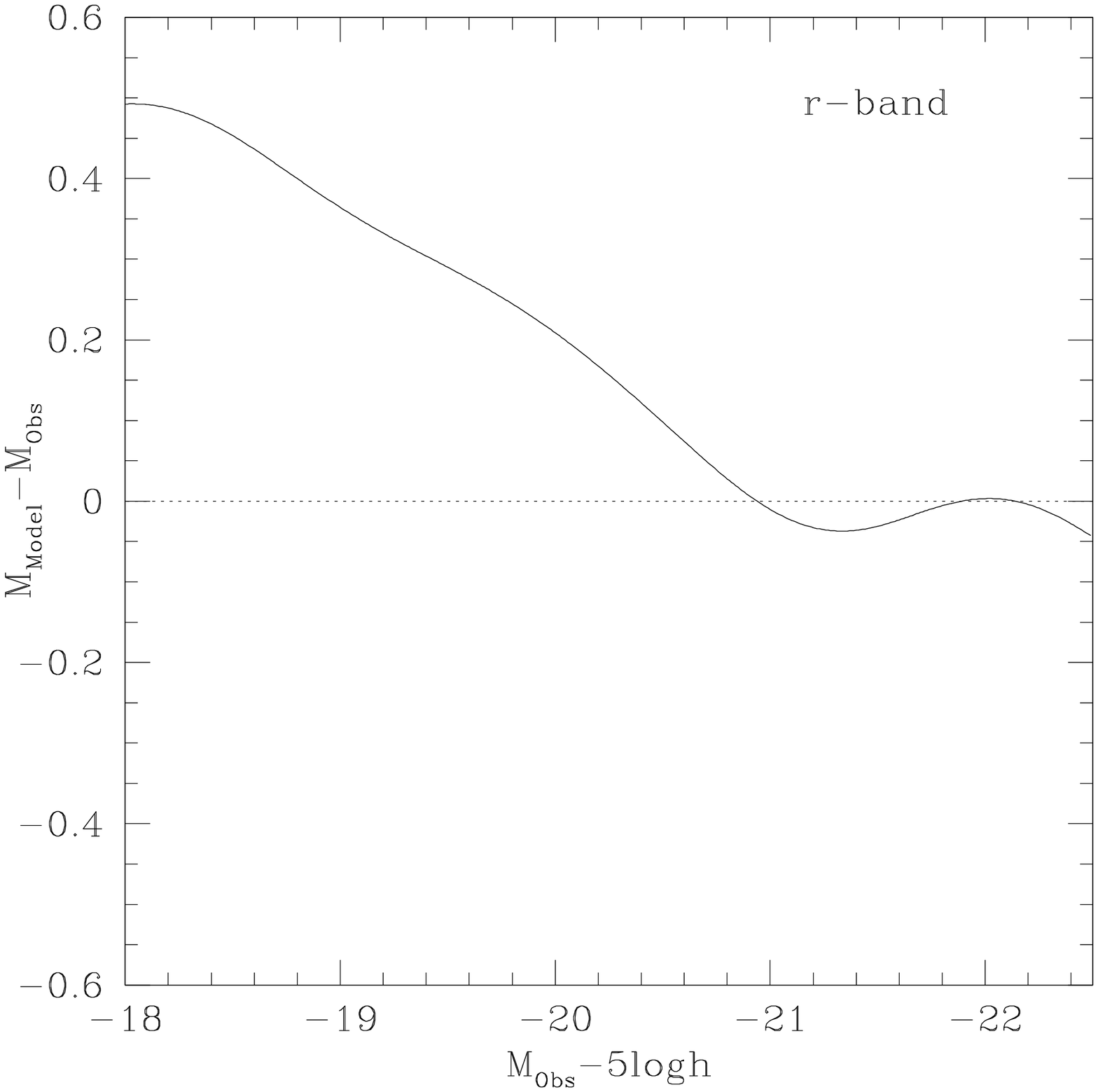,width=80mm}
\end{tabular}
\caption{The difference in model and observed absolute magnitudes for
which the number density of galaxies brighter than that magnitude is
the same as a function of observed absolute magnitude (solid
lines). Dotted lines show the relation if the model luminosity
function were to precisely match that observed. The left-hand panel
shows the relation in the ${\rm b_J}$-band, for which we use the
observed luminosity function of \protect\scite{norberg01}, while the
right-hand panel shows the relation for the r-band, for which we use
the observed luminosity function of \protect\scite{blanton01}.}
\label{fig:map}
\end{figure*}

It is important to note that the model of \scite{benson02} does not
produce a galaxy luminosity function which agrees precisely with that
observed (see their Fig.~9). As described in \scite{benson02b} we
modify the model of \scite{benson02} to use the \scite{jenkins01} dark
matter halo mass function (since this is a better match to the mass
function found in the N-body simulations that we use here than the
Press-Schechter mass function) and, to preserve reasonable agreement
with the observed luminosity functions, we adjust the star formation
timescale and mass-to-light ratios in the model.  The resulting model
luminosity function is not a perfect match to that observed. Many of
the void statistics employed in this work are sensitive to the number
density of points (i.e. galaxies).  To most fairly compare theory and
observation, we construct predicted samples of galaxies that match the
number density of the observed samples. Because of the difficulty in
exactly matching the shape of the luminosity function, matching the
number density leads to small offsets in the absolute magnitude limits
of the theoretical and observed samples, typically $|\Delta M|\lsim
0.4$. Therefore, we provide in Fig.~\ref{fig:map} a mapping between
observed and model absolute magnitudes which produce the same total
number density of brighter galaxies. (This corresponds to the results
which would be obtained if we were able to produce a good model
luminosity function by changing the luminosities of model galaxies
without changing the ranking of their luminosities.) Furthermore,
Table~\ref{tb:nmag} lists the number density of galaxies brighter than
a given absolute magnitude cut, for all cuts used in this
work. Throughout the remainder of this paper we will refer exclusively
to model magnitudes.

\begin{table}
\caption{The number densities of galaxies brighter than a given
absolute magnitude limit in our model. Values are given for all
magnitude cuts used in this work. We use model magnitudes exclusively
in this table.}
\label{tb:nmag}
\begin{center}
\begin{tabular}{ccc}
\hline
 & \multicolumn{2}{c}{$n(>M)/h^3$Mpc$^{-3}$} \\
$M-5\log h$ & 2dFGRS (b$_{\rm J}$-band) & SDSS (r-band) \\
\hline
$-18.0$ & $1.98 \times 10^{-2}$ & --- \\
$-18.5$ & $1.37 \times 10^{-2}$ & --- \\
$-19.5$ & $5.39 \times 10^{-3}$ & $2.00 \times 10^{-2}$ \\
$-20.0$ & $2.82 \times 10^{-3}$ & $1.37 \times 10^{-2}$ \\
$-20.5$ & $1.16 \times 10^{-3}$ & $9.02 \times 10^{-3}$ \\
$-21.0$ & $2.83 \times 10^{-4}$ & $2.92 \times 10^{-3}$ \\
$-21.5$ & $3.71 \times 10^{-5}$ & $1.22 \times 10^{-3}$ \\
$-22.0$ & $1.59 \times 10^{-5}$ & $2.37 \times 10^{-4}$ \\
$-22.5$ & $1.43 \times 10^{-5}$ & $3.63 \times 10^{-5}$ \\
$-23.0$ & ---                   & $6.61 \times 10^{-6}$ \\
\hline
\end{tabular}
\end{center}
\end{table}

We now describe the analyses which we will apply to our mock galaxy
catalogues.

\subsection{Void and Underdense Probability Functions}

The void probability function (VPF) for a distribution of points
(e.g. galaxies), as defined by \scite{white79}, is simply the
probability that a randomly placed sphere of radius $R$ will contain
no points within it. This statistic is sensitive to the presence of
voids in the distribution, and also has a particularly simple relation
to the halo occupation distributions considered by
\scite{benson01}. (In the notation of \scite{benson01}, the VPF is
just $S(0)$ when the halo distribution function $Q$ is computed for
spheres of radius $R$.) We compute the VPF by placing a large number of
such spheres in the simulation volumes and computing the fraction
which contain no galaxies. This process is repeated for a range of
sphere radii. We compute this statistic for samples of galaxies, dark
matter particles and randomly distributed particles at the same number
density as the galaxy sample in question. For a random set of points
the VPF has a particularly simple form:
\begin{equation}
P_0(R) = \exp (-4 \pi R^3 n /3),
\end{equation}
where $n$ is the mean number density of the points. We will also
consider briefly the underdense probability function (UPF) as defined
by \scite{vogeley94}. This statistic is defined as the probability
that the mean density of a randomly placed sphere of radius $R$ will
be below 20\% of the global mean density.  Unlike the VPF, the UPF
does not vary with the mean density of the sample, with the exception
of the small effect of galaxy discreteness on the threshold density.

\subsection{Nearest Neighbour Distance Distributions}

\scite{peebles01} has proposed the distribution of nearest neighbour
distances as a useful statistic to quantify to what degree one type of
galaxy respects the voids defined by another type. In this analysis we
define one sample of ``ordinary galaxies'' and use these to define the
locations of voids (we choose the name ``ordinary'' to distinguish
these from the ``wall galaxies'' categorized by the {\sc void finder}
algorithm of \S\ref{sec:vfa}). We then select a second set of
galaxies, which we refer to as ``test galaxies''. These two samples
can be chosen in any way we see fit, the idea being to choose test
galaxies which we suspect may populate the voids defined by the
ordinary galaxies. In this work we will use very simple criteria,
choosing ordinary galaxies to be those brighter than some particular
luminosity, and test galaxies to be a sample of fainter galaxies. We
compute the distance from each test galaxy to its nearest ordinary
galaxy, $D_{\rm to}$, and construct the distribution function of these
distances. Since $D_{\rm to}$ will depend on the number density of
ordinary galaxies we also compute the distribution function of $D_{\rm
oo}$, the distance from each ordinary galaxy to its nearest ordinary
galaxy.

By comparing the two distributions we can assess to what extent test
galaxies populate the voids defined by ordinary galaxies. For example,
if the distribution of $D_{\rm to}$ exceeds that of $D_{\rm oo}$ at
large distances we can infer that the test galaxies are populating the
voids.

\subsection{The Void Finder Algorithm}
\label{sec:vfa}

To study voids more directly we will apply the {\sc void finder}
algorithm of Hoyle \& Vogeley (2002; see also El-Ad \& Piran 1997) to
locate individual voids in our simulations and then proceed to
describe in detail below their properties and the properties of the
galaxies residing within them. The {\sc void finder} algorithm, as
used in this work, operates on a sample of galaxies and consists of
three steps, which we list below and then proceed to describe in
outline:
\begin{enumerate}
\item Categorise each galaxy in the sample as a wall galaxy or a void
galaxy.
\item Bin the wall galaxies into cells of a cubic grid.
\item Beginning from the centre of each empty grid cell, grow the
largest possible sphere containing no wall galaxies.
\item Find overlaps between maximal spheres and determine the set of unique voids by computing the overlap of these spheres.
\end{enumerate}
The void finder algorithm allows for some galaxies --- known as ``void
galaxies'' --- to lie within voids. These galaxies are those which
have three or fewer neighbours in a sphere of radius equal to $\langle
d\rangle+\frac{3}{2}\sigma_{\rm d}$ centred on it (where $\langle
d\rangle$ is the mean distance to the third nearest neighbour and
$\sigma_{\rm d}$ is the standard deviation of this distance). Using
this criteria, approximately 10\% of the galaxies are void galaxies
(consistent with the fraction found in observational samples by
\pcite{hv02}). Note that not all of these galaxies will in fact lie
within voids.

Wall galaxies (i.e. all non-void galaxies) are then assigned to cells
of a cubic grid.  From each empty grid cell, a maximal sphere
(i.e. the largest possible sphere which contains no wall galaxies) is
grown. These are referred to as ``holes''. As a single void will
typically contain more than one empty cell, there is some redundancy,
i.e. a single void will typically contain many holes. To determine
which of the holes are voids, we order the list of holes by size. The
largest hole is a void. Subsequent holes are only voids if they do not
overlap in volume with a previously detected void by more than
10\%. This choice of overlap fraction is somewhat arbitrary but is
discussed in greater detail in \scite{hv02}. In \scite{hv02} the
volumes of voids were enhanced by merging with smaller holes with
which they overlapped. In the present work we choose to consider only
the maximal sphere of each void, and so this volume enhancement
process is not carried out.

The result of this procedure is a list of voids in the simulation
volume, each of which is given a position (the centre of the sphere),
and the sphere radius. We use these lists to determine the
distribution of void sizes and the run of density with radius in
voids. By cross-referencing with our various galaxy catalogues we also
construct lists of galaxies residing within each void, which we use to
explore differences between the properties of void galaxies and wall
galaxies.

The catalogue of voids will become incomplete below some particular
radius due to the finite resolution of the computational grid. The
finite grid may also introduce some systematic bias in the sizes of
voids. We discuss incompleteness and bias separately below. For a grid
with cells of length $L$, the {\sc void finder} algorithm is
guaranteed to find all voids with radii larger than $\sqrt{3}L$. We
choose to use a grid of $64^3$ cells, which results in void catalogues
complete for radii in excess of $6.5$ and $15.7h^{-1}$Mpc for GIF and
$512^3$ simulations respectively. For smaller radii we expect the
catalogue to become gradually incomplete. Due to the nature of the
{\sc void finder} algorithm it is impossible to characterize the
incompleteness analytically and so we will make no use of the results
for voids of smaller radii. However, to approximately characterize the
incompleteness we applied the {\sc void finder} algorithm to one
galaxy sample using a $96^3$ element grid. We find that the results
using the $64^3$ grid are more than 90\% complete down to radii of
approximately 70\% of the nominal completeness radius. Below this the
completeness drops very rapidly.  Using the higher resolution grid
also demonstrates a small systematic bias in the recovered void
sizes. We find that, even for voids well above the completeness limit,
the higher resolution grid results in typical void sizes increasing by
around 4\%. This is due to the larger number of starting points for
the growth of holes resulting in a better chance of finding the true
maximal hole which can fit within a region of the sample (the effect
is clearly small so has little consequence for the results presented
in this work, but for the most accurate comparison of theory and
observations the two should be analysed using a grid of the same
resolution).

\section{Statistics of Voids}
\label{sec:statistics}

In this section we present results for the statistics of voids using
the analysis described in \S\ref{sec:analysis}.

\subsection{Void Probability Function}

\begin{figure*}
\begin{center}
\hspace{5mm}\psfig{file=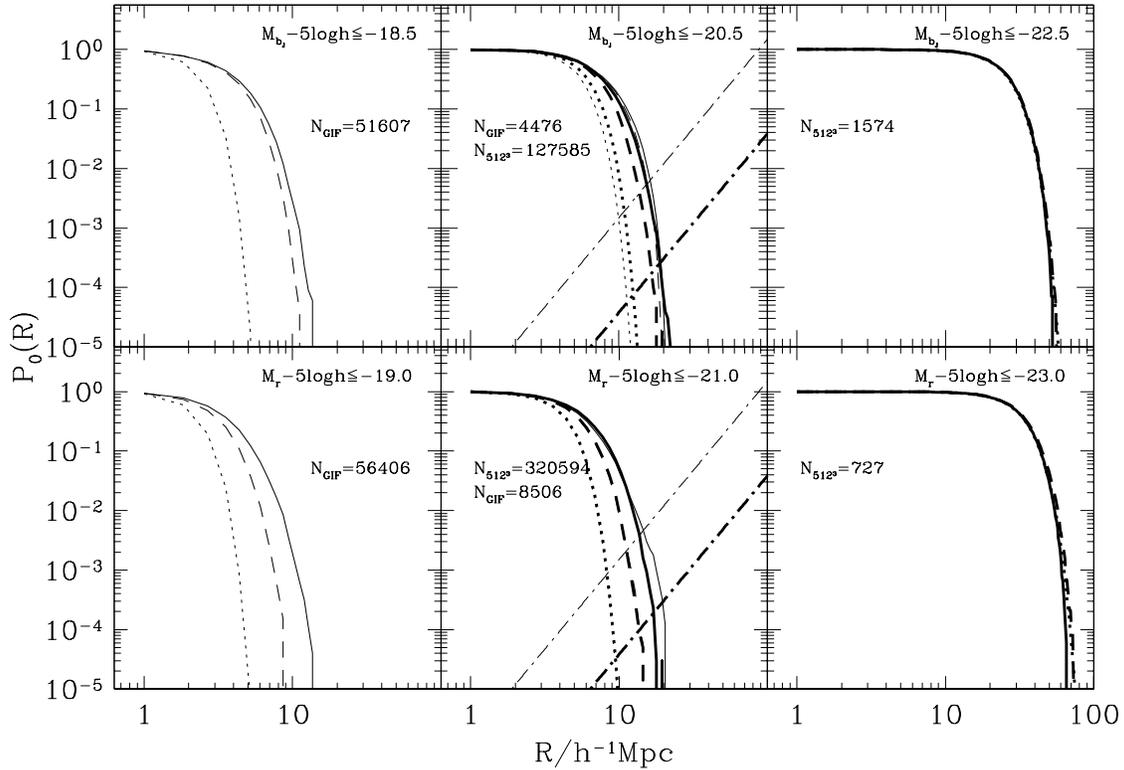,width=150mm,bbllx=5mm,bblly=132mm,bburx=190mm,bbury=265mm,clip=}
\end{center}
\caption{Void probability functions (VPF) for galaxies selected by
their b$_{\rm J}$ and r-band luminosities as appropriate to the 2dFGRS
(upper panels) and SDSS (lower panels) respectively. The absolute
magnitude selection is shown in each panel. In each panel solid lines
show the galaxy VPF, while dashed and dotted lines show the VPF of
dark matter and a random distribution of points with the same number
density as the galaxies. Thin lines show results from the GIF
simulation while heavy lines show results from the $512^3$
simulation. For the $M_{\rm b_{\rm J}}-5\log h\leq-20.5$ and $M_{\rm
r}-5\log h\leq-21$ samples we compare results from the two simulations
and find good agreement (the diagonal dot-dashed lines in these panels
indicate the point at which the average number of voids of given
radius in each simulation equals unity). The number of galaxies in
each sample is indicated in the panels.}
\label{fig:VPFboth}
\end{figure*}

In Fig.~\ref{fig:VPFboth} we show VPFs for galaxies selected by their
b$_{\rm J}$ and r-band magnitudes (upper and lower panels
respectively), with individual panels showing results for galaxies of
different luminosities. For fainter samples we use the GIF simulation
(shown by thin lines), while for the brighter samples we use the
$512^3$ simulation (shown by heavy lines) to obtain better
statistics. Where there is overlap between the two samples the
agreement is very good (the small horizontal offsets in the results
are due to slight differences in galaxy number density at fixed
absolute magnitude in the two simulations which arise due to noise in
the number of massive halos found in the GIF simulation). We indicate
in the relevant panels the regions for which the VPF is expected to be
determined accurately in each simulation. Furthermore, comparing
results from the GIF and GIF-II confirms that the VPF is determined to
high accuracy on these scales.

We compare the VPFs of our galaxies (solid lines) to those of dark
matter catalogues (dashed lines) and a random distribution of points
having the same number density (dotted lines). Several interesting
features are immediately obvious. Firstly, both dark matter and
galaxies contain many more large voids than a random sample of
points. This is not surprising of course given that both dark matter
and galaxies are clustered. Secondly, the VPF for large $R$
(e.g. $r\gsim 5$Mpc/h) is much higher for galaxy catalogues than for
dark matter catalogues. This indicates that galaxy catalogues contain
more large voids than the dark matter catalogues. This is a
consequence of galaxy bias --- few dark matter halos form in voids,
and those that do are typically of low mass and so almost never form a
galaxy bright enough to meet our selection criteria. Thus, although
low density regions always contain some dark matter they frequently
contain no galaxies.

\begin{figure}
\psfig{file=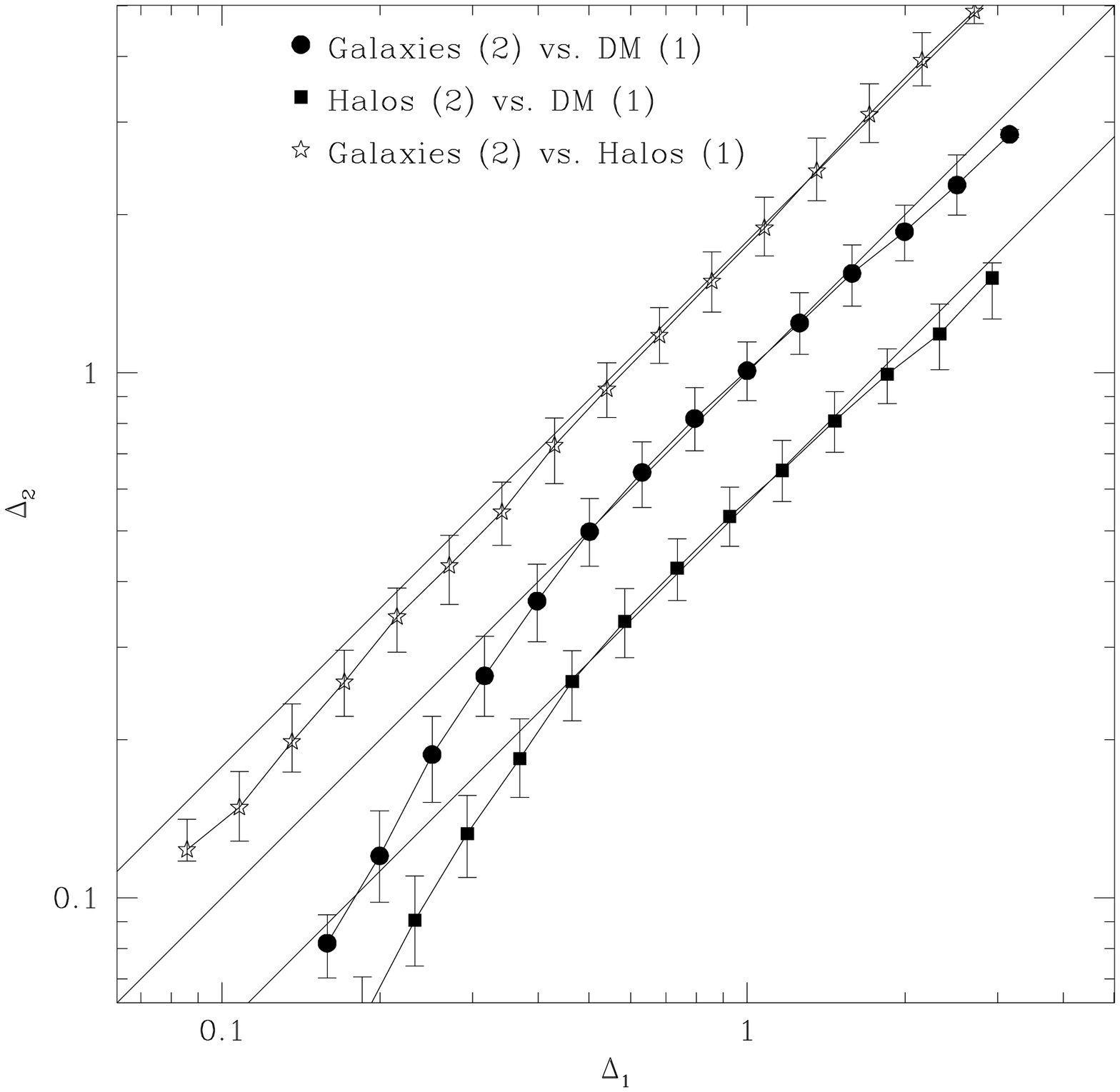,width=80mm}
\caption{The biasing relations between galaxies, halos and dark matter
density contrasts (the density contrasts $\Delta_1$ and $\Delta_2$
refer to different populations for each symbol type as indicated in
the figure labels). Results are shown for the GIF simulation, from
which samples of halos more massive than $10^{12}h^{-1}M_\odot$ and
galaxies brighter than $M_{\rm b_J}-5\log h=-19.5$ were extracted. The
distributions were smoothed with a Gaussian filter of scale
$10h^{-1}$Mpc. Points connected by lines show the median biasing
relations, with errorbars enclosing 80\% of the scatter in the
relation. Diagonal solid lines indicate an unbiased relation. Circles
show the relation between galaxies and dark matter, squares that
between halos and dark matter, and stars that between galaxies and
halos. The latter two relations are offset by $\pm 0.25$ in the
vertical direction for clarity.}
\label{fig:bias}
\end{figure}

We can address this point in more detail by examining the bias
relation between galaxies, dark matter and halos as shown in
Fig.~\ref{fig:bias}. We caution that we are using the term ``bias'' in
its most general sense --- namely as any difference between the
spatial distributions of galaxies and dark matter. It is possible that
a sample of galaxies that are biased in this sense may be unbiased in
terms of some particular statistic (e.g. the two-point correlation
function). For the purposes of this example we examine our $M_{\rm
b_J}-5\log h\leq-19.5$ galaxy sample, and a sample of dark matter
halos more massive than $10^{12}h^{-1}M_\odot$ (galaxies in this
sample are almost never found in lower mass halos, and most halos in
this mass range host at least one galaxy of this luminosity). The
distributions of galaxies, halos and dark matter in the GIF simulation
are smoothed with a Gaussian filter with scale $10h^{-1}$Mpc in order
to examine the bias on scales comparable to those of voids. The filled
circles in Fig.~\ref{fig:bias} show the median galaxy density
contrast\footnote{The density normalized to the mean density of the
sample in question, i.e. $\Delta = \rho/\overline{\rho}$ where $\rho$
is the local density and $\overline{\rho}$ is the mean density.}  as a
function of dark matter density contrast, with the errorbars giving an
indication of the scatter in the relation. The diagonal solid line
indicates the result for an unbiased population. We caution that due
to the small number of the highest density regions these results
become unreliable at large density contrasts. In regions of average
dark matter density the galaxy distribution is approximately
unbiased. In low density regions (i.e. voids) we see that galaxies are
biased, that is, their density contrast is typically lower than that
of the dark matter\footnote{Here we define `bias' as the ratio of
galaxy and dark matter overdensities, $b=\delta_{\rm g}/\delta_{\rm
DM}$, where $\delta= \rho/\overline{\rho}-1 = \Delta-1$. In underdense
regions, a 'biased' galaxy distribution therefore results in a lower
density contrast for galaxies than for dark matter.}. This effect is
responsible for the differences seen in galaxy and dark matter
VPFs. The filled squares show the bias relation for dark matter halos
(note that the points are offset for clarity). Clearly, the halos show
a similar biasing relation to the galaxies, specifically becoming more
underdense than the dark matter in low density regions. This result is
predicted by the extended Press-Schechter theory. This halo bias
explains most of the galaxy bias, but is not the whole effect. The
stars in Fig.~\ref{fig:bias} show the biasing relation between
galaxies and halos. Clearly these two populations are almost unbiased
relative to each other, particularly in high density regions. However,
in low density regions the galaxies are slightly more underdense than
halos. This relative bias is a consequence of galaxy formation, which
means that the lower mass halos in our sample (preferentially located
in the lower density regions) have a lower probability to host a
sufficiently bright galaxy. The combination of halo/dark matter and
galaxy/halo biases act constructively to produce the net galaxy/dark
matter bias, resulting in larger voids in the galaxy distribution than
in that of the dark matter. Note that this may not be the whole
story. We have probed the bias on one particular scale only. In fact,
as we will show in \S\ref{sec:dens}, galaxies are relatively more
underdense in the centres of voids than near the edges, which will
enhance the number of large galaxy voids relative to the number of
dark matter voids.

This bias is clearly visible in the galaxy and dark matter
distributions shown by Benson et al. (2001, their Fig.~1). Finally, it
is evident in Fig.~\ref{fig:VPFboth} that the difference between
galaxies, dark matter and random points is lessened for the brighter
(and so sparser) samples. This is, of course, just a reflection of the
lower number density of the brighter samples, which results in the
Poisson sampling of the underlying density field becoming the dominant
means of producing voids. These qualitative trends are seen for both
b$_{\rm J}$ and r-selected samples, and agree with those found by
\scite{kns97} using a similar technique.

\begin{figure}
\psfig{file=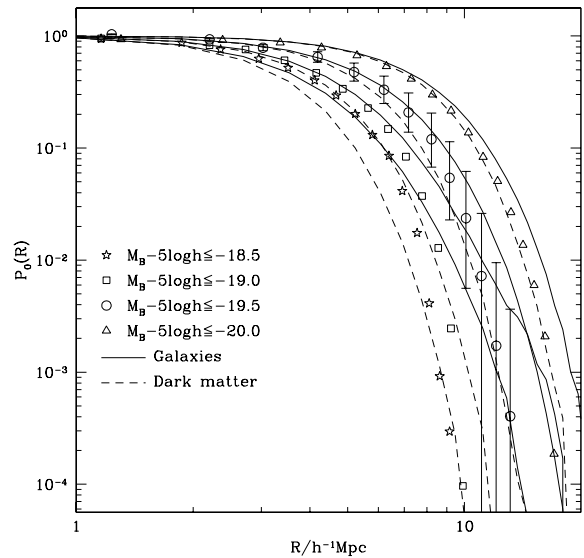,width=80mm}
\caption{The void probability functions for volume limited samples
drawn from the combined CfA-1 and CfA-2 galaxy redshift surveys
\protect\cite{vogeley94} are shown by symbols (see figure for
key). Error bars are shown for the $M_{\rm B}-5\log h\leq-19.5$ sample
only and include contributions due to the finite volume of the survey
and the uncertainty in the mean density due to fluctuations on scales
larger than the survey. For each observational sample we show the
predicted VPF for a sample of model galaxies with the same number
density (solid lines) and an equivalent sample of dark matter (dashed
lines). The luminosity corresponding to each line increases
monotonically from left to right. The model predictions are computed
from the $512^3$ simulation for the two brightest samples, and from
the GIF simulation for the others.}
\label{fig:VPFvog}
\end{figure}

\scite{vogeley94} measured the VPF for volume limited samples in the
Center for Astrophysics (CfA) surveys. In Fig.~\ref{fig:VPFvog} we
show their results for the combined CfA-1 and CfA-2 surveys for four
absolute magnitude limits. To compare our model to this data we
construct galaxy and dark matter catalogues with the same number
density as the samples analysed by \scite{vogeley94}. The resulting
model VPFs for galaxies and dark matter are shown in
Fig.~\ref{fig:VPFvog} as solid and dashed lines respectively. Note
that the magnitudes shown in this plot are the observational
magnitudes (i.e. not model magnitudes as referred to in the rest of
this paper).

\begin{figure}
\psfig{file=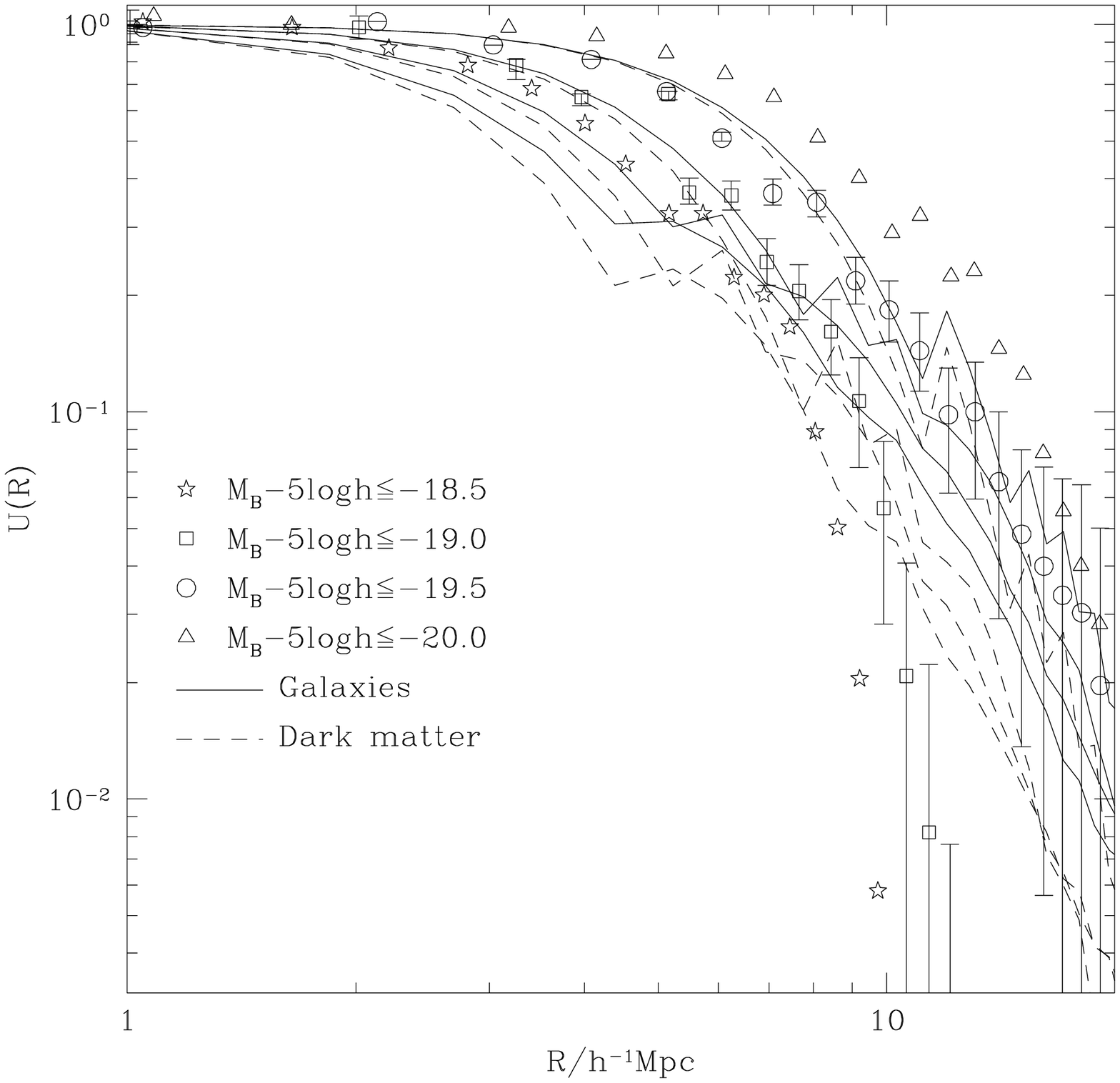,width=80mm}
\caption{The underdense probability functions for volume limited
samples drawn from the combined CfA-1 and CfA-2 galaxy redshift
surveys \protect\cite{vogeley94} are shown by symbols (see figure for
key). Error bars are shown for the $M_{\rm B}-5\log h\leq-19.5$ and
$-20.0$ samples only and are due to the finite volume of the
survey. For each observational sample we show the predicted UPF for a
sample of model galaxies with the same number density (solid lines)
and an equivalent sample of dark matter (dashed lines). The luminosity
corresponding to each line increases monotonically from bottom to top
(seen most clearly for $R$'s of a few $h^{-1}$Mpc). The model
predictions are computed from the $512^3$ simulation for the two
brightest samples, and from the GIF simulation for the others.}
\label{fig:UPFvog}
\end{figure}

This comparison of theory and observations shows an intriguing
dichotomy. For $R\lsim 8h^{-1}$Mpc our galaxy samples provide a good
match to the observed VPFs, while the dark matter samples produce
systematically low VPFs on these scales. On larger scales the
situation is reversed, with our model galaxy samples overpredicting
the observed VPFs, while dark matter samples agree rather well with
the data. The brightest sample is an exception, with the dark matter
sample agreeing best with the data on all scales. We must be cautious
however not to over-interpret the significance of these results. The
uncertainties in the observational determinations are shown in the
$M_{\rm B}-5\log h\leq -19.5$ sample. These include contributions from
both the finite number of independent volumes in the survey and the
uncertainty in the mean density due to the fluctuations on scales
larger than the survey. It should be kept in mind that the data points
in the VPF (both observational and theoretical versions) are
correlated. Given the current rather large observational
uncertainties, both the model galaxy and dark matter samples are
consistent with the data. Nevertheless, future observations could in
principle distinguish between the predictions for dark matter and for
galaxies.

In Fig.~\ref{fig:UPFvog} we show the UPF for the same observational
and model samples (again using model magnitudes). As the UPF is
independent of the number density we expect the results for dark
matter (dashed lines) to be identical in regions where shot noise is
unimportant. ``Shot noise'' here refers to the fluctuations in local
density caused by sparse sampling the underlying density field with a
finite number of galaxies. A secondary effect, due to the discreteness
in the number of galaxies which satisfy the density threshold, is
responsible for the sharp discontinuities in the UPF. As expected, the
three densest samples converge beyond about $6h^{-1}$Mpc (the sparsest
sample is still significantly affected by shot noise beyond this
radius). The galaxies show a bias relative to the dark matter,
producing a larger number of large radius underdense regions (the
effect is less visible for the sparsest sample where shot noise is
still a significant contribution on the scales probed here). On small
scales the model galaxies produce a UPF lower than that observed. On
large ($R\gsim 10h^{-1}$Mpc) scales the two brightest samples are in
good agreement with the observations, but the fainter samples
seriously overpredict the observed UPF. \scite{cen00} compared the UPF
for galaxies in a hydrodynamic simulation of a $\Lambda$CDM universe
with the observations of \scite{vogeley94}. Galaxies in their
simulation are also consistent with the observational data on scales
larger than $10h^{-1}$Mpc.

\subsection{Nearest Neighbour Distribution}

\begin{figure*}
\begin{center}
\hspace{5mm}\psfig{file=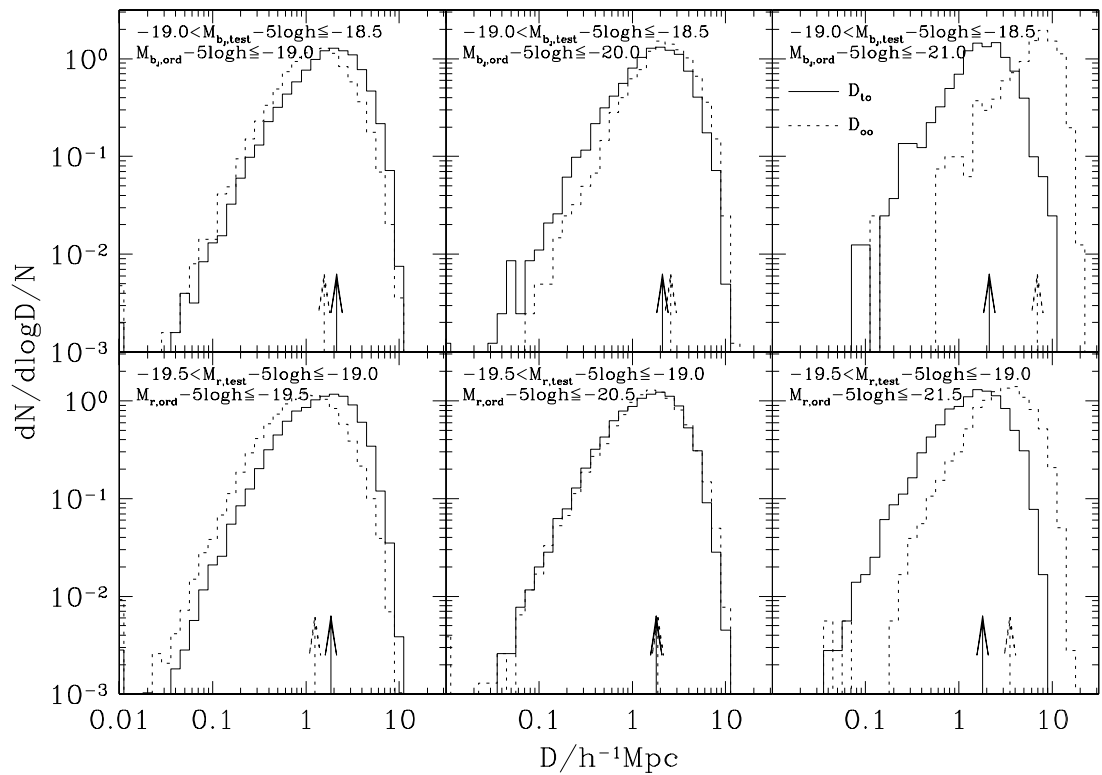,width=150mm,bbllx=5mm,bblly=132mm,bburx=190mm,bbury=265mm,clip=}
\end{center}
\caption{Distributions of distance to nearest neighbour for galaxy
samples selected by their b$_{\rm J}$ and r-band luminosity (as
appropriate to the 2dFGRS and SDSS respectively). In each panel we
indicate the absolute magnitude criteria used to select ``ordinary
galaxies'' and that used to select ``test galaxies''. The solid
histograms show the distribution of distances from each test galaxy to
the nearest ordinary galaxy, $D_{\rm to}$. The dotted histograms show
the distribution of distances from each ordinary galaxy to the nearest
ordinary galaxy, $D_{\rm oo}$, for comparison. Arrows indicate the
mean separation for each histogram. All results are computed using the
GIF simulation.}
\label{fig:NNboth}
\end{figure*}

In Fig.~\ref{fig:NNboth} we show distributions of nearest neighbour
distances for b$_{\rm J}$ and r-band selected galaxy samples (upper
and lower panels respectively). In each panel we use a sample of
bright ``ordinary'' galaxies (brighter than magnitude $M_{\rm ord}$ as
listed in each panel) to define the voids. We then use a fainter
sample (with magnitudes $M_{\rm test}$ in the range specified in each
panel) as test galaxies. The solid histograms show the distribution of
distances from each test galaxy to the nearest ordinary galaxy,
$D_{\rm to}$, while the dotted histogram shows the distribution of
distances from each ordinary galaxy to the nearest ordinary galaxy,
$D_{\rm oo}$, as a reference.

The peak of the $D_{\rm oo}$ distribution clearly shifts to larger
distances for the brighter samples of ordinary galaxies, as a
consequence of their lower number density (this is partially offset by
the stronger clustering of these galaxies, but this is a rather weak
effect) --- the vertical arrows in each panel show the mean separation
for each distribution. The $D_{\rm to}$ distribution for the fainter
ordinary samples is shifted to larger distances relative to that of
$D_{\rm oo}$. This indicates that the fainter test samples here do
begin to fill in the voids defined by the ordinary galaxies, with the
typical $D_{\rm to}$ being up to 50\% larger than the typical $D_{\rm
oo}$. For the brightest samples of ordinary galaxies we see an
opposite effect --- the $D_{\rm to}$ distribution peaks at smaller
distances than that of $D_{\rm oo}$. These bright ordinary galaxies
nearly always dwell at the centres of quite massive dark matter
halos. As such, two such ordinary galaxies are almost never found
within the same halo. Such an effect might not occur if we were able
to use significantly fainter samples of ordinary galaxies. A large
number of the faint test galaxies on the other hand are satellite
galaxies in the halos of the bright ordinary galaxies. As such, the
test galaxies typically live much closer to a bright ordinary galaxy
than do other bright ordinary galaxies. This is a crucial point in the
modelling of the nearest neighbour distribution which is missed in
calculations using halo centres as proxies for galaxies, and must be
considered when comparing observations with theory.

\scite{mathis02} performed a similar analysis, measuring the distances
from several samples of galaxies to their nearest bright spiral
galaxy. They found that very blue galaxies were the best candidate for
a population filling the voids, but none of their samples showed
evidence for a homogeneously distributed component. \scite{mathis02}
also noted that their results were in broad agreement with those of
\scite{peebles01}. Here, we have chosen to simply present
comprehensive predictions from our model which can be compared with
well-defined and large observational samples within the next few
years.

\section{Properties of Voids}
\label{sec:props}

\begin{figure}
\begin{center}
\psfig{file=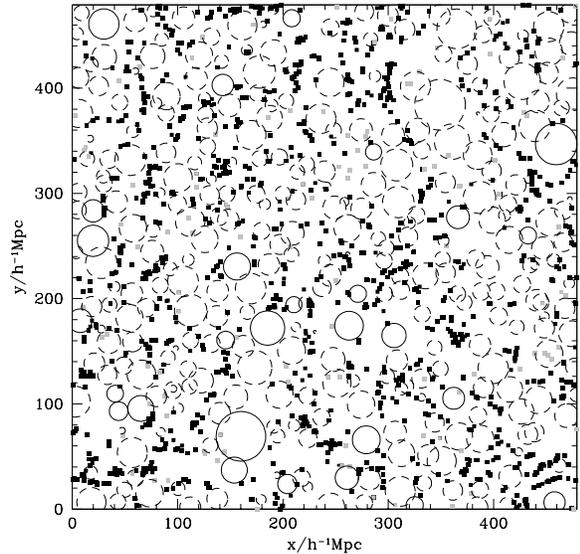,width=80mm}
\end{center}
\caption{The distribution of voids in a $2 h^{-1}$Mpc thick slice
through the $512^3$ simulation. The positions of ``wall galaxies'' (in
this specific case selected to be $M_{\rm b_{\rm J}}-5\log h \leq-20$
galaxies with more than three neighbours within a distance of
$6.9h^{-1}$Mpc) are shown by black squares. ``Void galaxies'' ---
those with three or fewer or neighbours within the above distance ---
are shown as light grey squares. Voids detected using the algorithm of
\protect\scite{hv02} are indicate by circles. Where the void centre
lies within the slice we show a solid-line circle with radius equal to
the radius of the void. Where the void centre lies outside of the
slice but some of the void overlaps the slice we show a dashed-line
circle with radius equal to the radius of the circle where void and
slice boundary intersect. Note that some wall galaxies appear to lie
inside of voids due to projection effects. In reality all voids are
devoid of wall galaxies.}
\label{fig:voidpic}
\end{figure}

We apply the {\sc void finder} algorithm to several samples of
galaxies and dark matter. The resulting void catalogues from the GIF
simulation are complete for voids with radii larger than
$6.5h^{-1}$Mpc, and for those larger than $15.6h^{-1}$Mpc in the
$512^3$ simulation. Figure~\ref{fig:voidpic} shows an example of the
distribution of these voids. It can be easily seen that voids
frequently contain some galaxies (note that the ``wall'' galaxies ---
shown as black dots in the figure --- are never really inside of
voids, they only appear to be due to projection effects), and that
small voids vastly outnumber larger voids. In the remainder of this
subsection we will quantify these points.

\subsection{Void Radius Distribution and Scaling Relations}

\begin{figure}
\begin{center}
\psfig{file=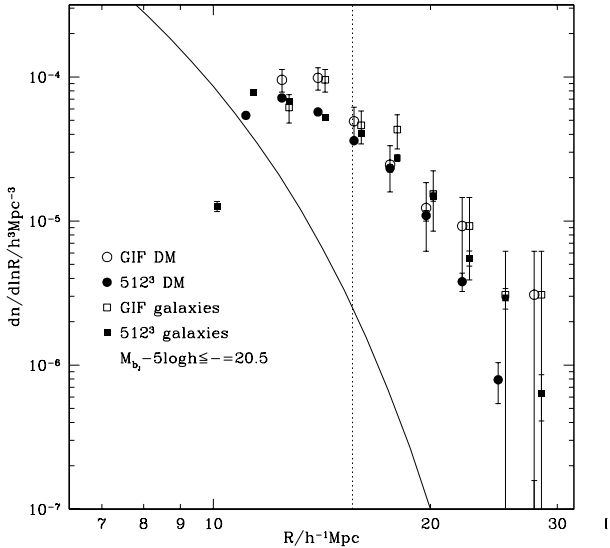,width=80mm}
\end{center}
\caption{The distribution of void radii. We plot the differential
number of voids of radius $R$ per unit volume as a function of
$R$. Points with errorbars show the results from our simulations,
while the solid line shows the prediction from the analytical model of
\protect\scite{sheth02}. Model results were obtained using samples of
galaxies brighter than $M_{\rm b_J}-5\log h=-20.5$ in the GIF and
$512^3$ simulation. We show estimates from the GIF and $512^3$
simulations and for dark and galaxy catalogues (symbol types as shown
in the figure key). The vertical dotted line indicates the smallest
radius for which the $512^3$ void catalogue is complete.}
\label{fig:voidsizescompare}
\end{figure}

The simplest property of voids which we can compute is the
distribution of their radii. Figure~\ref{fig:voidsizescompare} shows
void radii distributions for an $M_{\rm b_J}-5\log h\leq-20.5$ galaxy
sample and the corresponding dark matter sample in both the GIF and
$512^3$ simulations (see figure key for symbol types). Results from
the $512^3$ simulation are incomplete for radii smaller than that
indicated by the vertical dotted line, while those from
the GIF simulation are complete over the whole range of radii
shown. For larger voids the results from the GIF and $512^3$
simulations are in reasonable agreement.

Figure~\ref{fig:voidsizescompare} shows, as expected from the VPF
(Fig.~\ref{fig:VPFboth}), that there are more of the largest radii
voids in the galaxy distribution than in the dark matter
distribution. This is indicative that physical processes related to
galaxy formation as well as gravity help produce voids in the galaxy
distribution. At smaller radii however, the two distributions are very
similar. We predict a very rapid decline in the number density of
voids as their radii increase, with $\d n/\d \ln R$ falling by almost
three orders of magnitude over less than half an order of magnitude in
$R$. \scite{sheth02} has derived an analytical formulae for the
distribution of void sizes in the dark matter using arguments similar
to those employed by \scite{PS74}. The distribution predicted by
\scite{sheth02} is shown in Fig.~\ref{fig:voidsizescompare} by the
solid line. It should be noted that in Sheth's model a void is defined
as being a region in which the interior dark matter density contrast
is approximately $\Delta=\rho/\bar{\rho}=0.2$.

In contrast, our definition of voids (using the {\sc void finder})
yields voids that have even lower density contrast, typically
$\Delta=0.1$ or lower (especially for galaxy samples) as we will show
in \S\ref{sec:dens}. As such, we should not make a direct comparison
between our results and those of \scite{sheth02}. (In particular, the
analytic model of \scite{sheth02} predicts a unique distribution,
whereas our predicted distributions must necessarily depend on the
density of points in our galaxy samples.) For this particular sample,
the voids predicted by \scite{sheth02} are of comparable size to those
found by the {\sc void finder} algorithm (typically being only 40\%
and 50\% smaller than voids in our dark matter and galaxy catalogues
respectively). More interestingly, the rapid decline in void abundance
with radius seen in our model agrees well with that predicted by
\scite{sheth02}, who demonstrates that this is due to the fact that
the underdense regions of the Universe from which these voids form are
exponentially rare (assuming a Gaussian distribution of initial
density contrasts).

\begin{figure*}
\begin{center}
\psfig{file=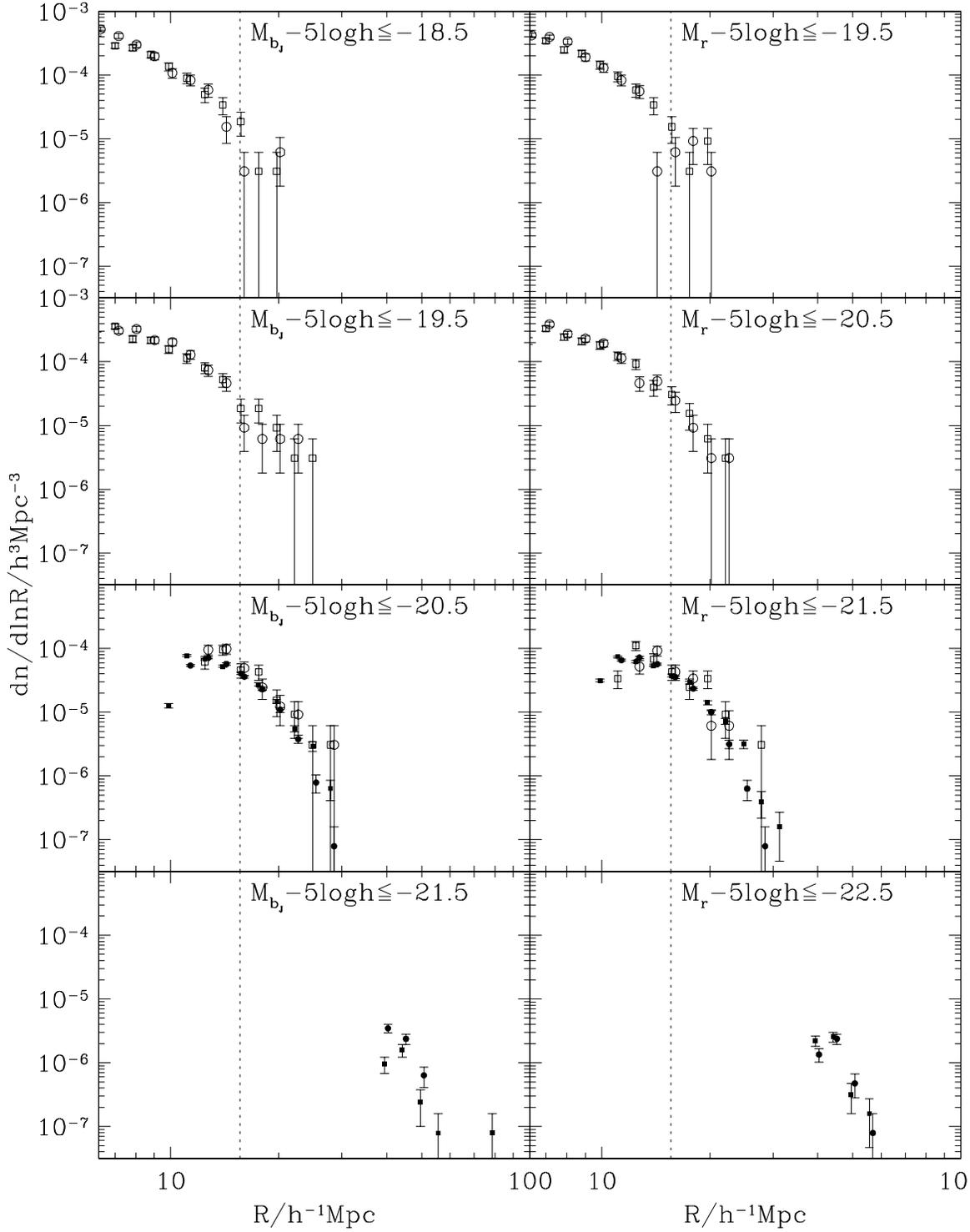,width=160mm,bbllx=0mm,bblly=30mm,bburx=187mm,bbury=265mm,clip=}
\end{center}
\caption{The distribution of void radii. We plot the differential
number of voids of radius $R$ per unit volume as a function of
$R$. Points with errorbars show the results from our simulations. Left
hand panels show samples selected by their b$_{\rm J}$-band magnitude,
while right-hand panels show samples selected by their r-band
magnitude (the magnitude selection is shown in each panel). Open and
filled symbols show results from the GIF and $512^3$ simulations
respectively. Squares and circles show results for galaxies and dark
matter respectively. The vertical dotted line shows the completeness
limit for the $512^3$ simulation. The GIF simulation is complete for
the whole range of radii shown.}
\label{fig:voidsizes}
\end{figure*}

In Fig.~\ref{fig:voidsizes} we show void radii distributions for
galaxy samples with different magnitude limits and selected in both
b$_{\rm J}$ and r-bands. As expected, for brighter (and so sparser)
samples the distribution shifts to larger sizes. The void radii
distributions show evidence of a peak (especially visible in the
$M_{\rm b_J}-5\log h\leq-20$ and $\leq -22$ panels), indicating a
characteristic size for voids. Finding voids using a higher resolution
grid does not change the position of this peak, indicating that it is
a real feature and not an artifact due to the limited resolution of
the computational grid. The limited dynamic range of the current
simulations allows us to say that the position of the peak moves to
larger radii as the sample of galaxies becomes brighter/sparser, but
prevents us from making any more quantitative conclusions (although we
{\it can} quantify the scaling of median void size with galaxy
luminosity as we will show below). The presence and changing position
of this peak is caused by the percolation of small voids into larger
ones, a process which depends on the mean density of the sample.

Note that while the GIF simulation appears to contain systematically
more large voids per unit volume than the $512^3$ simulation (e.g. see
the third row in Fig.~\ref{fig:voidsizes}) the difference is only
marginally significant. Assuming Poisson counting statistics the GIF
simulation exceeds the void abundance in the $512^3$ simulation by
$\lsim 2\sigma$ for these samples. We have also computed void radii
distributions using the GIF-II simulation (described in
\S\ref{sec:analysis}). We find that the Poisson errors are a reasonable
representation of the differences between the results from the GIF and
GIF-II simulations, while the GIF-II results are somewhat closer to
those of the $512^3$ simulation.

\scite{sheth02} also predicts the clustering properties of voids
(specifically their bias on large scales and defined in terms of the
two-point correlation function). Even with the large volume of our
simulations measuring void clustering is quite difficult. On scales
comparable to the void radii we find a strong anti-correlation since
the {\sc void finder} algorithm allows voids to overlap by at most
10\% in volume. On larger scales we typically find that voids have a
close to uniform distribution. However, given the small amplitude of
both void and dark matter correlation functions on these scales it is
impossible to make strong statements about the void bias, or how it
scales with void radius, from our current simulations.

\begin{figure*}
\begin{center}
\begin{tabular}{cc}
\psfig{file=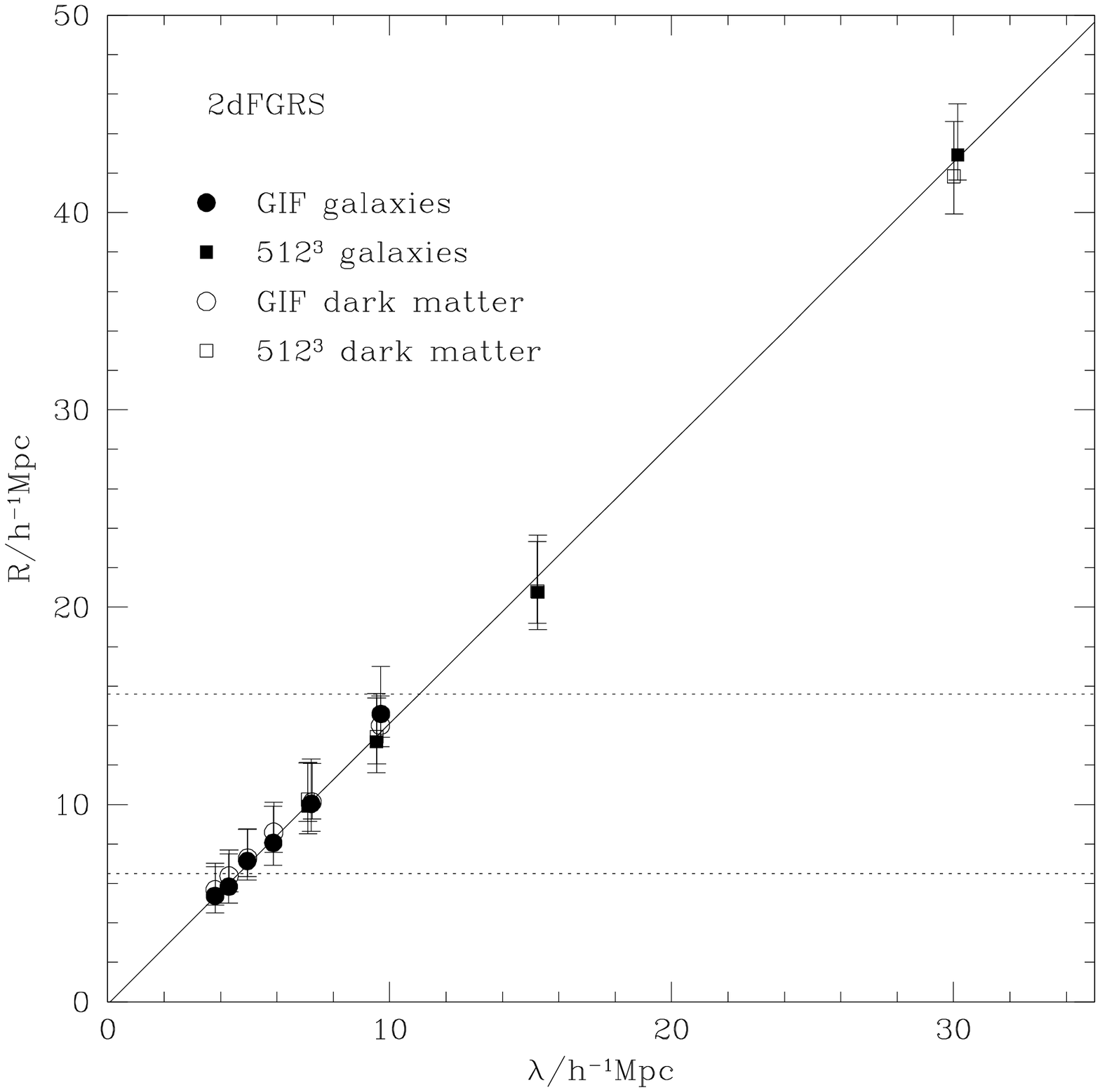,width=80mm} & \psfig{file=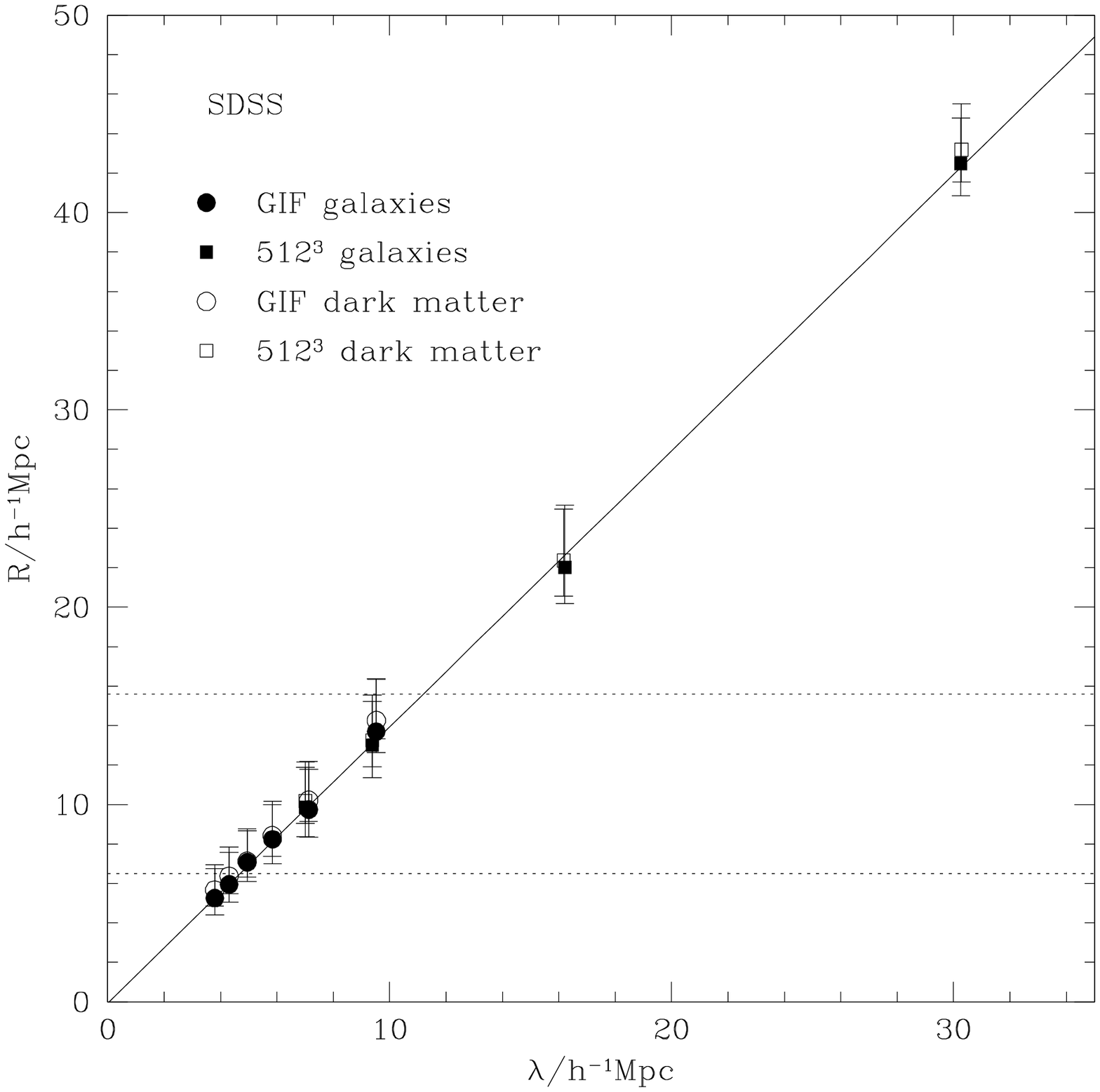,width=80mm}
\end{tabular}
\end{center}
\caption{The median void radius, $R$, as a function of mean
inter-galaxy separation, $\lambda$. The left-hand panel shows results
for galaxies selected by their b$_{\rm J}$-band magnitude (as
appropriate to the 2dFGRS), with the right-hand panel showing results
for galaxies selected by their r-band magnitude (as appropriate to the
SDSS). Open symbols show results for dark matter catalogues, while
filled symbols show results for galaxy catalogues. Circles and squares
show results from the GIF and $512^3$ simulations
respectively. Errorbars enclose 50\% of the distribution of void
radii. Horizontal dotted lines indicate the completeness limit for
each simulation. The solid line indicates the best fit linear relation
to the galaxy results, including only those points for which more than
75\% of the voids have radii greater than the completeness limit.}
\label{fig:voidscaling}
\end{figure*}

\scite{armu} show that voids in redshift surveys and in mock
galaxy catalogues built from CDM simulations obey a simple scaling
relation, such that the median void radius, $R$, obeys
$R=R_0+\nu\lambda$, where $\lambda$ is the mean galaxy separation of
the sample in question and $R_0$ and $\nu$ are parameters. In
Fig.~\ref{fig:voidscaling} we show the $R$--$\lambda$ relation for
voids in our simulations. We compute $R$ for all the galaxy samples
using both the GIF and $512^3$ simulations. Note that there is little
difference seen in the median void size between galaxy and dark matter
samples. The differences in the void size distribution seen in
Fig.~\ref{fig:voidsizescompare} occur only for the largest voids,
which have very low number density and so have little impact on the
median void size. The parameters $R_0$ and $\nu$ are determined using
a least squares fit, including only those points for which 75\% of the
voids in the sample have radii in excess of the completeness limit for
the simulation in question. For the 2dFGRS sample we find $R_0=-0.11\pm
0.6$ and $\nu=1.42\pm0.04$, while for the SDSS sample we find
$R_0=-0.06\pm 0.2$ and $\nu=1.40\pm0.02$. Thus, the scaling relations
for our two samples are consistent within the errorbars. Our results
differ somewhat from those of \scite{armu}, who find higher values
of $\nu$, and larger values of $R_0$ for CDM mock galaxy catalogues
constructed using a simple biassing scheme and for the Las Campanas
Redshift Survey. This is not surprising given that we employ a
different void finding algorithm, but does serve to highlight the
importance of analyzing model and observations using the same
algorithm. Nevertheless, such a linear scaling relation with slope
$\nu>1$ is found in both void detection schemes, suggesting that it
may be a generic feature of the void distribution.

\subsection{Void Density Profiles}
\label{sec:dens}

\begin{figure*}
\begin{center}
\psfig{file=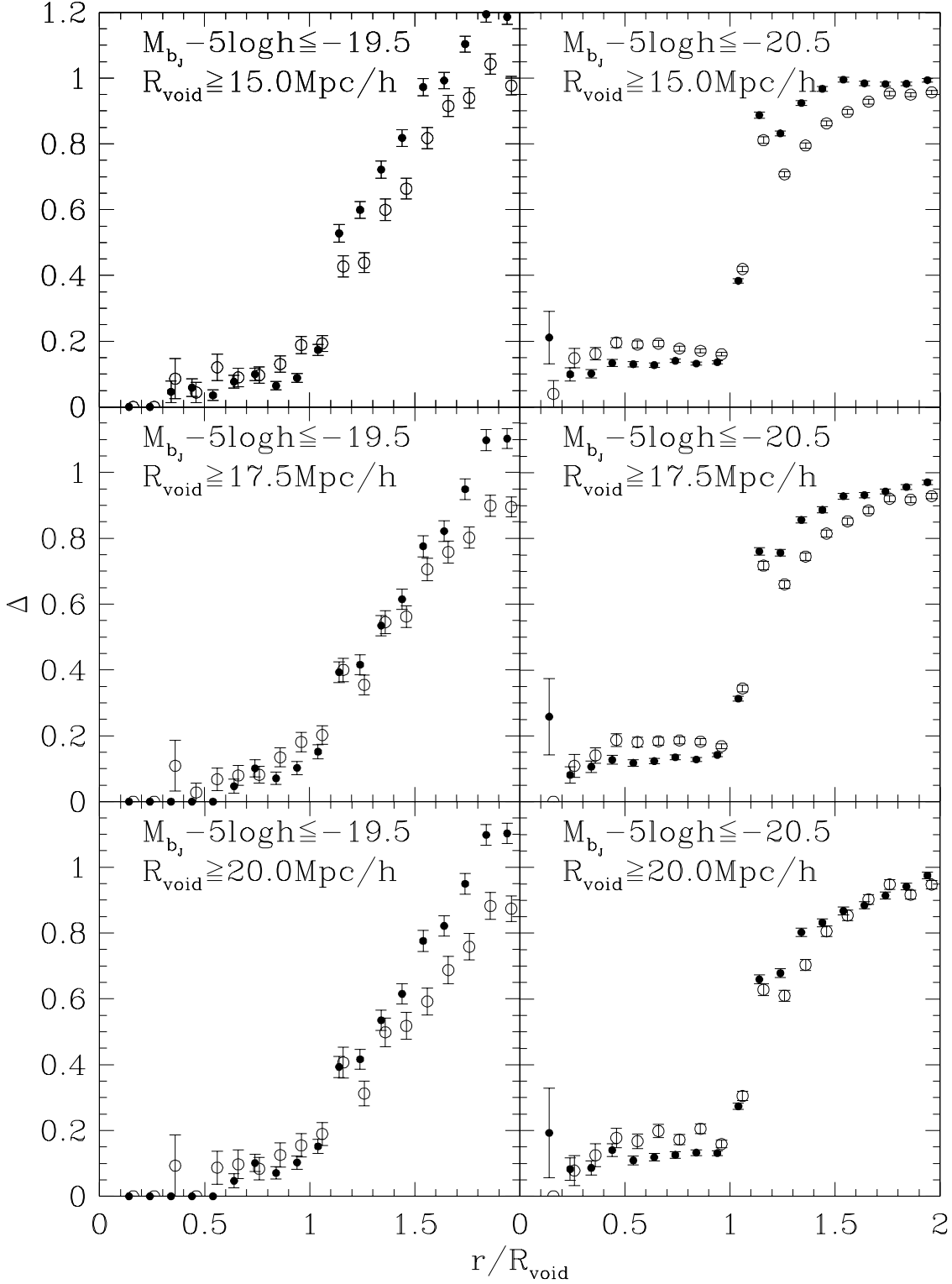,width=155mm,bbllx=0mm,bblly=17mm,bburx=187mm,bbury=265mm,clip=}
\end{center}
\caption{Mean density contrast as a function of radial position in
voids. We plot the mean density contrast as a function of scaled
radius (i.e. we express radius in each void in units of the void
radius) and averaged over all voids in our GIF and $512^3$ simulations
(left and right-hand panels respectively). Left and right-hand panels
show results for voids found in galaxy samples selected to have
$M_{\rm b_J}-5\log h\leq-19.5$ and $-20.5$ respectively. Upper, middle
and lower panels include those voids with radii larger than $15$,
$17.5$ and $20h^{-1}$Mpc respectively. Open circles show results for
dark matter catalogues, while filled circles show results for galaxy
catalogues.}
\label{fig:voidprofile}
\end{figure*}

In the {\sc void finder} algorithm voids need not be entirely devoid
of galaxies. Galaxies with few nearby neighbours are classified as
``void galaxies'' and \emph{may} be located within the subsequently
detected voids. To assess the prevalence of such galaxies, and the
larger scale structure surrounding voids, we have determined void
density contrast profiles from our simulations. To do this we
determine the number density of galaxies in concentric shells centred
on each void centre, scaling all length scales by the void radius to
allow us to compare voids of different sizes. We sum the resulting
number density profiles over all voids in a particular range of radii
and convert this into a density contrast profile. The results of this
procedure are shown in Fig.~\ref{fig:voidprofile}.

We show results for both galaxy and dark matter catalogues (filled and
open points respectively), for a range of minimum void sizes and for
two different magnitude selections (as indicated in the panels). Voids
are highly underdense, galaxy voids more so than dark matter voids
(typically density contrasts are $0.1$ and $0.2$ respectively for the
$M_{\rm b_J}-5\log h\leq-20.5$ sample), again demonstrating the
biasing of galaxies relative to dark matter (this can be seen most
clearly in the upper left-hand panel of
Fig.~\ref{fig:voidprofile}). There is little variation in density
contrast within much of the void, but there is evidence for a decline
in density contrast in the very central regions of the voids. There is
a clear threshold corresponding to the edge of each void which occurs
close to $r/R_{\rm void}=1$, especially for the $M_{\rm b_J}-5\log
h\leq-20.5$ sample, indicating that the {\sc void finder} algorithm is
working successfully. Beyond the void radius we find the density
contrast increases rapidly (more rapidly for galaxies than dark
matter), but can remain below unity for a significant distance. Thus,
for our $M_{\rm b_J}-5\log h\leq-20.5$ sample the regions around voids
are, on average, still underdense even at twice the void radius. These
findings are in qualitative agreement with the void density profiles
reported by \scite{armu}, with the exception that our voids are
less dense (for both galaxies and dark matter), and their boundaries
are sharper (i.e. density contrast increases more rapidly beyond the
void radius). This is again a consequence of the differing algorithms
for defining voids in the two approaches, highlighting the importance
of analysing observations and theoretical models in identical ways.

\section{Properties of Void Galaxies}
\label{sec:vgals}

\begin{figure*}
\psfig{file=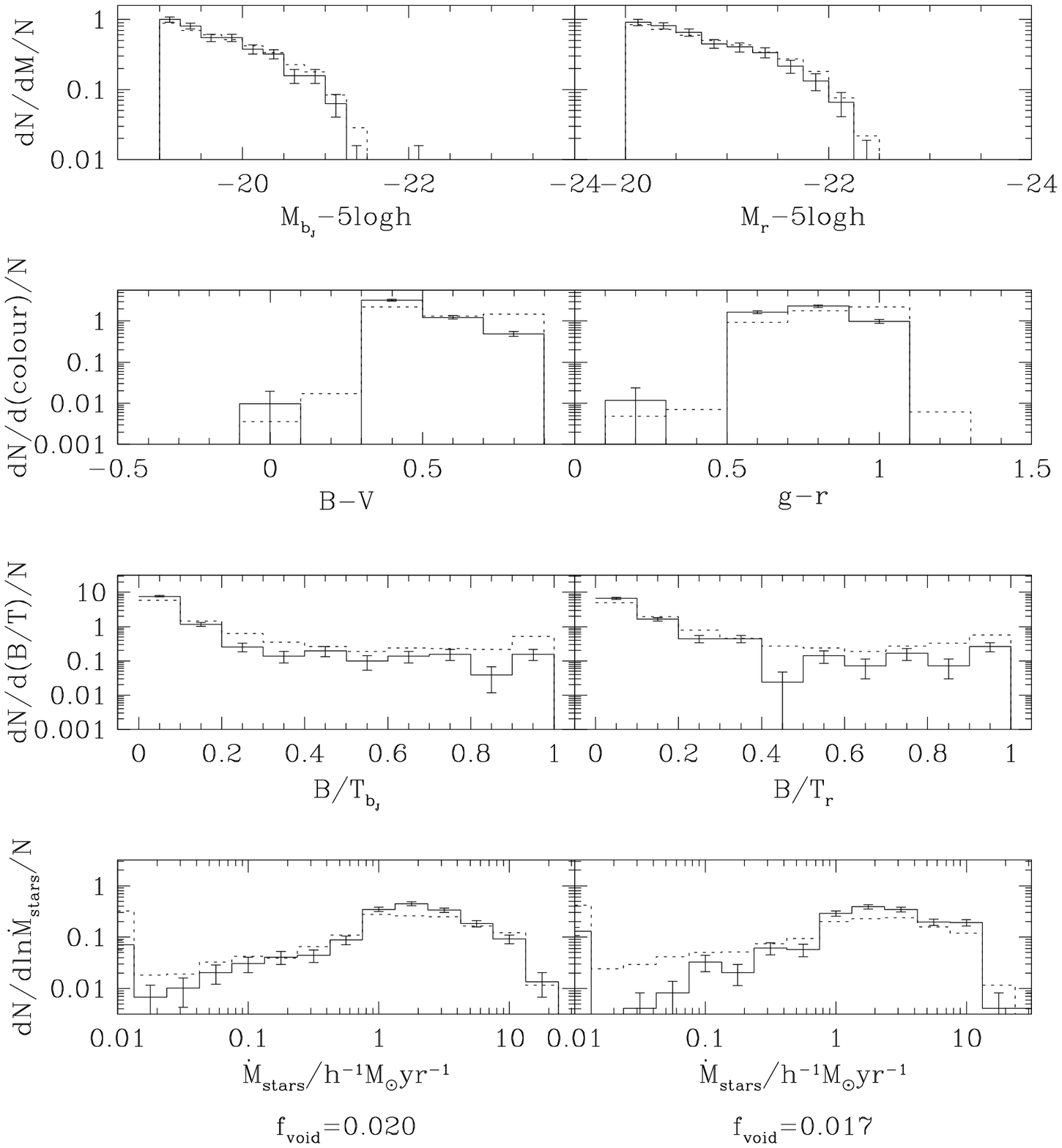,width=160mm,bbllx=5mm,bblly=10mm,bburx=190mm,bbury=215mm,clip=}
\caption{Normalized distributions of b$_{\rm J}$/r-band absolute
magnitude, B$-$V/g$-$r colour, bulge-to-total ratio (measured in
dust-extinguished b$_{\rm J}$-band light) and star formation rate (top
to bottom panels respectively) for 2dFGRS/SDSS samples for the general
population of galaxies (dotted histograms) and galaxies living inside
voids of radius 12.5Mpc/h or larger (solid histograms with error bars
indicating the Poisson variation). In all cases we limit the sample of
galaxies to those brighter than b$_{\rm J}-5\log h \leq -19$ and ${\rm
r}-5\log h\leq-20$ for 2dFGRS and SDSS samples respectively. The
fraction of galaxies in these samples which live in such voids,
$f_{\rm void}$, is shown at the bottom of the figure. All results are
from the GIF simulation.}
\label{fig:vgalsGIF}
\end{figure*}

In Fig.~\ref{fig:vgalsGIF} we show, for the GIF simulation, the
distributions of galaxy luminosities, colours, morphologies and star
formation rates for both the general population (dotted histograms)
and for galaxies living within voids (i.e. those galaxies lying within
the spherical voids found using the {\sc void finder}; solid
histograms). Note that the fraction of galaxies which are actually
located within voids (shown at the bottom of Fig.~\ref{fig:vgalsGIF})
is much lower than the fraction of galaxies classified as ``void
galaxies'' by the {\sc void finder} algorithm. A similar result is
found for the observational samples employed by \scite{hv02}. It is
clear that void galaxies have systematically different properties to
the general population---although the differences are typically rather
small (similar results are found in the $512^3$
simulation). Nevertheless, a Kolmogorov-Smirnov test indicates that
the void and field galaxy samples are inconsistent with being drawn
from a single underlying distribution (the probability being less than
6\% in all cases). Void galaxies are typically fainter and bluer, are
more disk-dominated and show higher star formation rates. All of these
differences can be traced to the difference in dark matter halo mass
functions in voids (indeed, in our model, this is the only possible
means of creating a difference between voids and the general
field). For example, in the field some fraction of galaxies fall into
clusters where their supply of fresh gas is ``strangled'' (i.e. their
diffuse halos of hot gas is removed by the ram pressure and tidal
forces due to the cluster), resulting in a cessation of star
formation. Since there are no clusters in voids this process is far
less important for void galaxies.

\begin{figure}
\psfig{file=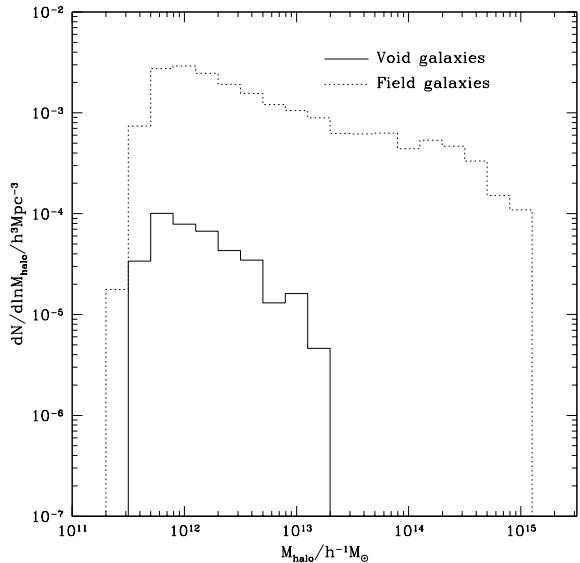,width=80mm}
\caption{The (galaxy number weighted-)dark matter halo mass function
for a sample of $M_{\rm b_J}-5\log h\leq-19$ galaxies from the GIF
simulation. The dotted line shows that of all such galaxies, while the
solid line shows the result for galaxies living inside voids of radius
$12.5h^{-1}$Mpc or larger.}
\label{fig:halomf}
\end{figure}

For the samples shown in Fig.~\ref{fig:vgalsGIF} the halo mass
function (for halos occupied by one of the galaxies) is shown in
Fig.~\ref{fig:halomf}. Clearly the halo mass function of void galaxies
(solid line) is shifted to much lower masses relative to that of the
field sample (dotted line). The median occupied halo mass is over 6
times lower for void galaxies than for field galaxies.  The majority
of these void galaxies live at the centre of their halo, whereas in
the general population we find a large fraction of the galaxies
existing as satellites in more massive halos.

\begin{figure}
\psfig{file=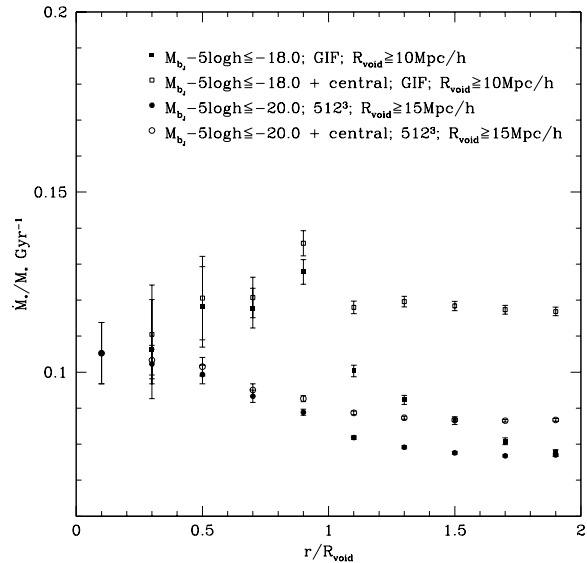,width=80mm}
\caption{The specific star formation rate (i.e. the star formation
rate per unit mass of stars) in model galaxies as a function of the
radius relative to their nearest void centre. The mean specific star
formation rate is plotted for two different luminosity selections
(squares and circles, as indicated in the figure) as a function of
normalized radius to the nearest void centre. Filled symbols include
all model galaxies in the specified luminosity range, while open
symbols include only those living at the centre of their halo (i.e. it
excludes satellite galaxies). Results are computed for voids larger
than $10$ and $15h^{-1}$Mpc for the faint and bright samples
respectively.}
\label{fig:voidSFR}
\end{figure}

We are also able to explore how the star formation rate of galaxies
varies with radius inside voids. In Fig.~\ref{fig:voidSFR} we show the
mean specific star formation rate (i.e. the star formation rate per
unit mass of stars) in model galaxies as a function of their
normalized distance from the nearest void centre, where we include
only voids larger than a particular radius (as indicated in the
figure). Results are shown for two b$_{\rm J}$-band selected samples
as indicated in the figure (the fainter one drawn from the GIF
simulation, the brighter one from the $512^3$ simulation). Results for
the full samples of galaxies are shown by the filled points. There is
a clear trend for higher specific star formation rates as we move
towards the centres of voids (for the fainter sample in particular,
the trend is seen out to twice the void radius). There are two
contributions to this effect. Firstly, outside of voids we find very
massive halos which contain large populations of satellite
galaxies. In our model these galaxies have lost their supply of fresh
gas, and so star formation is strongly suppressed in these
systems. Since such massive halos are very rare in voids the majority
of void galaxies are the dominant galaxy of their halo, so retain a
gas supply allowing higher rates of star formation. We can see the
contribution of this effect by repeating this calculation using only
central galaxies (i.e. those with a continued gas supply), as shown by
the open symbols in Fig.~\ref{fig:voidSFR}. For the fainter sample in
particular this can be seen to be the dominant contributor to the
trend. A secondary contributor to the trend is that, even for galaxies
with a continued gas supply, specific star formation rates tend to be
higher for lower mass galaxies in our model. This weaker trend is
visible in the open symbols in Fig.~\ref{fig:voidSFR}.

Larger differences \emph{may} occur for fainter samples of galaxies,
below the resolution limit of our simulations. If faint field galaxies
exist mostly as satellites in massive groups and clusters (which are
not located in voids) then significant differences between field and
void samples could occur. If, on the other hand, faint field galaxies
exist mostly in lower mass halos (where the differences between field
and void halo mass functions are much smaller) we would expect the
differences to be smaller also. Benson et al. (in preparation) examine
the relative contributions of satellite and non-satellite galaxies to
the galaxy luminosity function, finding that satellites become the
dominant contribution faintwards of $M_{\rm B}-5\log h\approx -16.5$,
suggesting that we would need to probe over two magnitudes fainter
than currently possible to see significant differences between void
and field galaxies.

\section{Discussion}
\label{sec:discuss}

We have presented detailed theoretical predictions for a wide range of
statistics related to voids in the distribution of galaxies, and for
properties of galaxies living within those voids. These predictions
have been specifically aimed at the 2dFGRS and SDSS, which, due to
their large size, should provide accurate measures of these
statistics. We have focussed on simple selection criteria, namely
simple cuts in luminosity, which we believe will permit the most
robust comparison between theory and observations.

The properties of voids and void galaxies are potentially a strong
constraint on models of structure and galaxy formation. However, we
have shown that many of the properties of galaxy and dark matter voids
differ significantly, indicating (as we have shown) that galaxy bias
as well as gravitational instability is important for the formation of
galaxy voids. Although our understanding of galaxy bias has progressed
greatly in recent years we cannot be certain that our understanding is
complete. Where suitable data already exist (specifically the void and
underdense probability functions analysis applied to the CfA surveys
by \pcite{vogeley94}) we find that the model is consistent with the
data, but that the current observational sample is too small to probe
the signal of bias expected from our model. This situation should be
rectified when this analysis is repeated on larger redshift surveys.

As has been shown in previous works on galaxy clustering using
semi-analytic and N-body techniques, the results are rather robust to
changes in model parameters \cite{cluster1}. If we compute void
statistics such as the VPF, nearest neighbour distributions or void
size distributions at a fixed number density then our predictions are
unaffected by changes in most model parameters (e.g. the strength of
supernovae feedback, the star formation timescales in galaxies
etc.). This results from the fact that the main effect of changing
these parameters is to change galaxy luminosities without
significantly changing the ranking of galaxy luminosity.  Predictions
are changed by parameters in the model which do change the ranking of
luminosities. For example, significantly increasing or decreasing the
rates of galaxy mergers can make strong differences in some of the
statistics considered here. (Note however, that with the improved
merging model of \scite{benson02} we no longer have the freedom to
adjust merger rates in our model.) The properties of void galaxies
(e.g. Fig.~\ref{fig:vgalsGIF}) \emph{are} affected by changes in model
parameters. As the differences between void and wall galaxies seen in
Fig.~\ref{fig:vgalsGIF} are so small we choose not to explore the
dependencies of these properties on model parameters in this work.

The models employed in our analysis include the effects of
``photoionization suppression'' as described by \scite{benson02}. This
feedback mechanism has the potential to strongly alter the properties
of voids in the galaxy distribution. Halos in voids are of lower mass
on average than in higher density regions. Since photoionization
suppression acts most effectively on low-mass halos it will cause
greatest suppression of galaxy formation in voids. (Furthermore,
although not included in our present modelling, reionization of the
Universe may begin in voids, enhancing the suppression in these
regions further.) For the lowest mass galaxies resolved in our current
N-body simulations the effects of photoionization suppression are
negligible (e.g. the VPF and nearest neighbour distributions for
galaxy samples of fixed number density are indistinguishable between
models with and without photoionization suppression). We may expect
however strong differences to show up in higher resolution
simulations.

\section*{Acknowledgements}

We acknowledge valuable conversations with Ravi Sheth and Tommaso
Treu. We thank Carlton Baugh, Shaun Cole, Carlos Frenk and Cedric
Lacey for permitting us to use the {\sc galform} semi-analytic model
of galaxy formation. The N-body simulations used in this work were
kindly made available by the Virgo Consortium. FH and MSV acknowledge
support from NSF grant AST-0071201 and the John Templeton
Foundation. FT was supported by the Caltech SURF program.


\begin{thebibliography}
\bibitem[Amendola et al. <1999>]{amendola99}Amendola~L., Baccigalupi~C., Gleiser~M., Occionero~F., 1999, New Astronomy, 3, 339
\bibitem[Arbabi-Bidgoli \& M\"uller <2002>]{armu}Arbabi-Bidgoli~S., M\"uller~V., 2002, MNRAS, 332, 205
\bibitem[Benson et al. <2000a>]{cluster1}Benson~A.~J., Cole~S., Frenk~C.~S., Baugh~C.~M., Lacey~C.~G., 2000a, MNRAS, 311, 793
\bibitem[Benson et al. <2000b>]{cluster2}Benson~A.~J., Baugh~C.~M., Cole~S., Frenk~C.~S., Lacey~C.~G., 2000b, MNRAS, 316, 10
\bibitem[Benson et al. <2001>]{cluster3}Benson~A.~J., Frenk~C.~S., Baugh~C.~M., Cole~S., Lacey~C.~G., 2001, MNRAS, 327, 1041
\bibitem[Benson <2001>]{benson01}Benson~A.~J., 2001, MNRAS, 325, 1039
\bibitem[Benson et al. <2002a>]{benson02}Benson~A.~J., Lacey~C.~G., Baugh~C.~M., Cole~S., Frenk~C.~S., 2002a, MNRAS, 333, 156
\bibitem[Benson et al. <2002b>]{benson02b}Benson~A.~J., Lacey~C.~G., Baugh~C.~M., Cole~S., Frenk~C.~S., 2002b, in preparation
\bibitem[Blanton et al. <2001>]{blanton01}Blanton~M.~R. et al., 2001, ApJ, 121, 2358
\bibitem[Cen \& Ostriker <2000>]{cen00}Cen~R., Ostriker~J.~P., 2000, ApJ, 538, 83
\bibitem[Cole et al. <2000>]{cole2000}Cole~S., Lacey~C.~G., Baugh~C.~M., Frenk~C.~S., 2000, MNRAS, 319, 168
\bibitem[Davis \& Geller <1976>]{davis76}Davis~M., Geller~M.~J., 1976, ApJ, 208, 1
\bibitem[Einasto et al. <1991>]{einasto91}Einasto~J., Einasto~M., Gramann~M., Saar~E., 1991, 248, 593
\bibitem[El-Ad \& Piran <1997>]{elad97}El-Ad~H., Piran~T., 1997, ApJ, 491, 421
\bibitem[Geller \& Huchra <1989>]{geller89}Geller~M.~J., Huchra~J.~P., 1989, Science, 246, 897
\bibitem[Ghigna et al. <1994>]{ghigna94}Ghigna~S., Borgani~S., Bonometto~S.~A., Guzzo~L., Klypin~A., Primack~J.~R., Giovanelli~R., Haynes~M.~P., 1994, ApJ, 437, 71
\bibitem[Ghigna et al. <1996>]{ghigna96}Ghigna~S., Bonometto~S.~A., Retzlaff~J., Gottl\"ober~S., Murante~G., ApJ, 469, 40
\bibitem[Gregory \& Thompson <1978>]{greg78}Gregory~S.~A., Thompson~L.~A., 1978, ApJ, 222, 784
\bibitem[Hoyle et al. <1999>]{hoyle99}Hoyle~F., Baugh~C.~M., Shanks~T., Ratcliffe~A., 1999, MNRAS, 309, 659
\bibitem[Hoyle \& Vogeley <2002>]{hv02}Hoyle~F., Vogeley~M.~S., 2002, ApJ, 566, 641
\bibitem[Jenkins et al. <2001>]{jenkins01}Jenkins~A., Frenk~C.~S., White~S.~D.~M., Colberg~J.~M., Cole~S., Evrard~A.~E., Couchman~H.~M.~P., Yoshida~N., 2001, MNRAS, 321, 372
\bibitem[Kauffmann, Nusser \& Steinmetz <1997>]{kns97}Kauffmann~G., Nusser~A., Steinmetz~M., 1997, MNRAS, 286, 795
\bibitem[Kauffmann et al. <1999>]{kauffmann99}Kauffmann~G., Colberg~J.~M., Diaferio~A., White~S.~D.~M., 1999, MNRAS, 303, 188
\bibitem[Kirshner et al. <1981>]{kirshner81}Kirshner~R.~P., Oemler~A.~Jr., Schechter~P.~L., Shectman~S.~A., 1981, ApJ, 248, L57
\bibitem[Little \& Weinberg <1994>]{little94}Little~B., Weinberg~D.~H., 1994, MNRAS, 267, 605
\bibitem[Mathis \& White <2002>]{mathis02}Mathis~H., White~S.~D.~M., 2002, submitted to MNRAS (astro-ph/0201193)
\bibitem[M\"uller et al. <2000>]{muller00}M\"uller~V., Arbabi-Bidgoli~S., Einasto~J., Tucker~D., 2000, MNRAS, 318, 280
\bibitem[Norberg et al. <2001a>]{norberg01a}Norberg~P. et al., 2001a, MNRAS, 328, 64
\bibitem[Norberg et al. <2002>]{norberg01}Norberg~P. et al., 2002, MNRAS, 332, 827
\bibitem[Ostriker \& Cowie <1981>]{ocow81}Ostriker~J.~P., Cowie~L.~L., 1981, ApJ, 243, L127
\bibitem[Peebles <2001>]{peebles01}Peebles~P.~J.~E., 2001, astro-ph/0101127
\bibitem[Press \& Schechter <1974>]{PS74}Press~W.~H., Schechter~P., 1974, ApH, 187, 425
\bibitem[Sheth <2002>]{sheth02}Sheth~R.~K., 2002, submitted to MNRAS
\bibitem[Verde et al. <2002>]{lverde}Verde~L. et al., 2002, submitted to MNRAS (astro-ph/0112161)
\bibitem[Vogeley et al. <1991>]{vogeley91}Vogeley~M.~S., Geller~M.~J., Huchra~J.~P., ApJ, 1991, 382, 44
\bibitem[Vogeley et al. <1994>]{vogeley94}Vogeley~M.~S., Geller~M.~J., Park~C., Huchra~J.~P., 1994, AJ, 108, 745
\bibitem[White <1979>]{white79}White~S.~D.~M., 1979, MNRAS, 186, 145
\bibitem[Zehavi et al. <2002>]{zehavi02}Zehavi~I. et al. (the SDSS Collaboration), ApJ, 571, 172
\bibitem[Zehavi et al. <2003>]{zehavi03}Zehavi~I. et al., in preparation
\end{thebibliography}
\end{document}